\documentclass[journal,twoside,twocolumn]{IEEEtran}
\usepackage{cite}
\usepackage[T1]{fontenc}
\usepackage{graphicx}
\usepackage{xcolor}
\usepackage{amssymb}
\usepackage{amsmath}
\usepackage{graphicx}
\usepackage{subfigure}
\usepackage{booktabs}
\usepackage{algorithm}
\usepackage{algorithmic}
\usepackage{amsthm}
\usepackage{multirow}
\usepackage{microtype}
\usepackage{balance}

\usepackage{array}
\usepackage{threeparttable}

\usepackage{amssymb}
\usepackage{amsfonts}
\usepackage{mathrsfs}
\usepackage{xspace}
\usepackage{bm}
\usepackage{upgreek}

\newcommand{\safemath}[2]{\newcommand{#1}{\ensuremath{#2}\xspace}}

\safemath{\bma}{\mathbf{a}}
\safemath{\bmb}{\mathbf{b}}
\safemath{\bmc}{\mathbf{c}}
\safemath{\bmd}{\mathbf{d}}
\safemath{\bme}{\mathbf{e}}
\safemath{\bmf}{\mathbf{f}}
\safemath{\bmg}{\mathbf{g}}
\safemath{\bmh}{\mathbf{h}}
\safemath{\bmi}{\mathbf{i}}
\safemath{\bmj}{\mathbf{j}}
\safemath{\bmk}{\mathbf{k}}
\safemath{\bml}{\mathbf{l}}
\safemath{\bmm}{\mathbf{m}}
\safemath{\bmn}{\mathbf{n}}
\safemath{\bmo}{\mathbf{o}}
\safemath{\bmp}{\mathbf{p}}
\safemath{\bmq}{\mathbf{q}}
\safemath{\bmr}{\mathbf{r}}
\safemath{\bms}{\mathbf{s}}
\safemath{\bmt}{\mathbf{t}}
\safemath{\bmu}{\mathbf{u}}
\safemath{\bmv}{\mathbf{v}}
\safemath{\bmw}{\mathbf{w}}
\safemath{\bmx}{\mathbf{x}}
\safemath{\bmy}{\mathbf{y}}
\safemath{\bmz}{\mathbf{z}}
\safemath{\bmzero}{\mathbf{0}}
\safemath{\bmone}{\mathbf{1}}

\bmdefine{\biad}{a}
\bmdefine{\bibd}{b}
\bmdefine{\bicd}{c}
\bmdefine{\bidd}{d}
\bmdefine{\bied}{e}
\bmdefine{\bifd}{f}
\bmdefine{\bigd}{g}
\bmdefine{\bihd}{h}
\bmdefine{\biid}{i}
\bmdefine{\bijd}{j}
\bmdefine{\bikd}{k}
\bmdefine{\bild}{l}
\bmdefine{\bimd}{m}
\bmdefine{\bind}{n}
\bmdefine{\biod}{o}
\bmdefine{\bipd}{p}
\bmdefine{\biqd}{q}
\bmdefine{\bird}{r}
\bmdefine{\bisd}{s}
\bmdefine{\bitd}{t}
\bmdefine{\biud}{u}
\bmdefine{\bivd}{v}
\bmdefine{\biwd}{w}
\bmdefine{\bixd}{x}
\bmdefine{\biyd}{y}
\bmdefine{\bizd}{z}

\bmdefine{\bixid}{\xi}
\bmdefine{\bilambdad}{\lambda}
\bmdefine{\bimud}{\mu}
\bmdefine{\bithetad}{\theta}
\bmdefine{\biphid}{\phi}
\bmdefine{\bideltad}{\delta}

\safemath{\bmia}{\biad}
\safemath{\bmib}{\bibd}
\safemath{\bmic}{\bicd}
\safemath{\bmid}{\bidd}
\safemath{\bmie}{\bied}
\safemath{\bmif}{\bifd}
\safemath{\bmig}{\bigd}
\safemath{\bmih}{\bihd}
\safemath{\bmii}{\biid}
\safemath{\bmij}{\bijd}
\safemath{\bmik}{\bikd}
\safemath{\bmil}{\bild}
\safemath{\bmim}{\bimd}
\safemath{\bmin}{\bind}
\safemath{\bmio}{\biod}
\safemath{\bmip}{\bipd}
\safemath{\bmiq}{\biqd}
\safemath{\bmir}{\bird}
\safemath{\bmis}{\bisd}
\safemath{\bmit}{\bitd}
\safemath{\bmiu}{\biud}
\safemath{\bmiv}{\bivd}
\safemath{\bmiw}{\biwd}
\safemath{\bmix}{\bixd}
\safemath{\bmiy}{\biyd}
\safemath{\bmiz}{\bizd}

\safemath{\bmxi}{\bixid}
\safemath{\bmlambda}{\bilambdad}
\safemath{\bmmu}{\bimud}
\safemath{\bmtheta}{\bithetad}
\safemath{\bmphi}{\biphid}
\safemath{\bmdelta}{\bideltad}

\safemath{\bA}{\mathbf{A}}
\safemath{\bB}{\mathbf{B}}
\safemath{\bC}{\mathbf{C}}
\safemath{\bD}{\mathbf{D}}
\safemath{\bE}{\mathbf{E}}
\safemath{\bF}{\mathbf{F}}
\safemath{\bG}{\mathbf{G}}
\safemath{\bH}{\mathbf{H}}
\safemath{\bI}{\mathbf{I}}
\safemath{\bJ}{\mathbf{J}}
\safemath{\bK}{\mathbf{K}}
\safemath{\bL}{\mathbf{L}}
\safemath{\bM}{\mathbf{M}}
\safemath{\bN}{\mathbf{N}}
\safemath{\bO}{\mathbf{O}}
\safemath{\bP}{\mathbf{P}}
\safemath{\bQ}{\mathbf{Q}}
\safemath{\bR}{\mathbf{R}}
\safemath{\bS}{\mathbf{S}}
\safemath{\bT}{\mathbf{T}}
\safemath{\bU}{\mathbf{U}}
\safemath{\bV}{\mathbf{V}}
\safemath{\bW}{\mathbf{W}}
\safemath{\bX}{\mathbf{X}}
\safemath{\bY}{\mathbf{Y}}
\safemath{\bZ}{\mathbf{Z}}

\safemath{\bZero}{\mathbf{0}}
\safemath{\bOne}{\mathbf{1}}
\safemath{\bDelta}{\mathbf{\Delta}}
\safemath{\bLambda}{\mathbf{\UpLambda}}
\safemath{\bPhi}{\mathbf{\Upphi}}
\safemath{\bSigma}{\mathbf{\Upsigma}}
\safemath{\bOmega}{\mathbf{\Upomega}}
\safemath{\bTheta}{\mathbf{\Uptheta}}

\bmdefine{\biAd}{A}
\bmdefine{\biBd}{B}
\bmdefine{\biCd}{C}
\bmdefine{\biDd}{D}
\bmdefine{\biEd}{E}
\bmdefine{\biFd}{F}
\bmdefine{\biGd}{G}
\bmdefine{\biHd}{H}
\bmdefine{\biId}{I}
\bmdefine{\biJd}{J}
\bmdefine{\biKd}{K}
\bmdefine{\biLd}{L}
\bmdefine{\biMd}{M}
\bmdefine{\biOd}{N}
\bmdefine{\biPd}{O}
\bmdefine{\biQd}{P}
\bmdefine{\biRd}{R}
\bmdefine{\biSd}{S}
\bmdefine{\biTd}{T}
\bmdefine{\biUd}{U}
\bmdefine{\biVd}{V}
\bmdefine{\biWd}{W}
\bmdefine{\biXd}{X}
\bmdefine{\biYd}{Y}
\bmdefine{\biZd}{Z}

\bmdefine{\biDelta}{\Delta}
\bmdefine{\biLambda}{\Lambda}
\bmdefine{\biPhi}{\Phi}
\bmdefine{\biSigma}{\Sigma}
\bmdefine{\biOmega}{\Omega}
\bmdefine{\biTheta}{\Theta}

\safemath{\bimA}{\biAd}
\safemath{\bimB}{\biBd}
\safemath{\bimC}{\biCd}
\safemath{\bimD}{\biDd}
\safemath{\bimE}{\biEd}
\safemath{\bimF}{\biFd}
\safemath{\bimG}{\biGd}
\safemath{\bimH}{\biHd}
\safemath{\bimI}{\biId}
\safemath{\bimJ}{\biJd}
\safemath{\bimK}{\biKd}
\safemath{\bimL}{\biLd}
\safemath{\bimM}{\biMd}
\safemath{\bimN}{\biNd}
\safemath{\bimO}{\biOd}
\safemath{\bimP}{\biPd}
\safemath{\bimQ}{\biQd}
\safemath{\bimR}{\biRd}
\safemath{\bimS}{\biSd}
\safemath{\bimT}{\biTd}
\safemath{\bimU}{\biUd}
\safemath{\bimV}{\biVd}
\safemath{\bimW}{\biWd}
\safemath{\bimX}{\biXd}
\safemath{\bimY}{\biYd}
\safemath{\bimZ}{\biZd}

\safemath{\bimDelta}{\biDelta}
\safemath{\bimLambda}{\biLambda}
\safemath{\bimPhi}{\biPhi}
\safemath{\bimSigma}{\biSigma}
\safemath{\bimOmega}{\biOmega}
\safemath{\bimTheta}{\biTheta}

\safemath{\setA}{\mathcal{A}}
\safemath{\setB}{\mathcal{B}}
\safemath{\setC}{\mathcal{C}}
\safemath{\setD}{\mathcal{D}}
\safemath{\setE}{\mathcal{E}}
\safemath{\setF}{\mathcal{F}}
\safemath{\setG}{\mathcal{G}}
\safemath{\setH}{\mathcal{H}}
\safemath{\setI}{\mathcal{I}}
\safemath{\setJ}{\mathcal{J}}
\safemath{\setK}{\mathcal{K}}
\safemath{\setL}{\mathcal{L}}
\safemath{\setM}{\mathcal{M}}
\safemath{\setN}{\mathcal{N}}
\safemath{\setO}{\mathcal{O}}
\safemath{\setP}{\mathcal{P}}
\safemath{\setQ}{\mathcal{Q}}
\safemath{\setR}{\mathcal{R}}
\safemath{\setS}{\mathcal{S}}
\safemath{\setT}{\mathcal{T}}
\safemath{\setU}{\mathcal{U}}
\safemath{\setV}{\mathcal{V}}
\safemath{\setW}{\mathcal{W}}
\safemath{\setX}{\mathcal{X}}
\safemath{\setY}{\mathcal{Y}}
\safemath{\setZ}{\mathcal{Z}}
\safemath{\emptySet}{\varnothing}

\safemath{\colA}{\mathscr{A}}
\safemath{\colB}{\mathscr{B}}
\safemath{\colC}{\mathscr{C}}
\safemath{\colD}{\mathscr{D}}
\safemath{\colE}{\mathscr{E}}
\safemath{\colF}{\mathscr{F}}
\safemath{\colG}{\mathscr{G}}
\safemath{\colH}{\mathscr{H}}
\safemath{\colI}{\mathscr{I}}
\safemath{\colJ}{\mathscr{J}}
\safemath{\colK}{\mathscr{K}}
\safemath{\colL}{\mathscr{L}}
\safemath{\colM}{\mathscr{M}}
\safemath{\colN}{\mathscr{N}}
\safemath{\colO}{\mathscr{O}}
\safemath{\colP}{\mathscr{P}}
\safemath{\colQ}{\mathscr{Q}}
\safemath{\colR}{\mathscr{R}}
\safemath{\colS}{\mathscr{S}}
\safemath{\colT}{\mathscr{T}}
\safemath{\colU}{\mathscr{U}}
\safemath{\colV}{\mathscr{V}}
\safemath{\colW}{\mathscr{W}}
\safemath{\colX}{\mathscr{X}}
\safemath{\colY}{\mathscr{Y}}
\safemath{\colZ}{\mathscr{Z}}

\safemath{\opA}{\mathbb{A}}
\safemath{\opB}{\mathbb{B}}
\safemath{\opC}{\mathbb{C}}
\safemath{\opD}{\mathbb{D}}
\safemath{\opE}{\mathbb{E}}
\safemath{\opF}{\mathbb{F}}
\safemath{\opG}{\mathbb{G}}
\safemath{\opH}{\mathbb{H}}
\safemath{\opI}{\mathbb{I}}
\safemath{\opJ}{\mathbb{J}}
\safemath{\opK}{\mathbb{K}}
\safemath{\opL}{\mathbb{L}}
\safemath{\opM}{\mathbb{M}}
\safemath{\opN}{\mathbb{N}}
\safemath{\opO}{\mathbb{O}}
\safemath{\opP}{\mathbb{P}}
\safemath{\opQ}{\mathbb{Q}}
\safemath{\opR}{\mathbb{R}}
\safemath{\opS}{\mathbb{S}}
\safemath{\opT}{\mathbb{T}}
\safemath{\opU}{\mathbb{U}}
\safemath{\opV}{\mathbb{V}}
\safemath{\opW}{\mathbb{W}}
\safemath{\opX}{\mathbb{X}}
\safemath{\opY}{\mathbb{Y}}
\safemath{\opZ}{\mathbb{Z}}
\safemath{\opZero}{\mathbb{O}}
\safemath{\identityop}{\opI}

\safemath{\veca}{\bma}
\safemath{\vecb}{\bmb}
\safemath{\vecc}{\bmc}
\safemath{\vecd}{\bmd}
\safemath{\vece}{\bme}
\safemath{\vecf}{\bmf}
\safemath{\vecg}{\bmg}
\safemath{\vech}{\bmh}
\safemath{\veci}{\bmi}
\safemath{\vecj}{\bmj}
\safemath{\veck}{\bmk}
\safemath{\vecl}{\bml}
\safemath{\vecm}{\bmm}
\safemath{\vecn}{\bmn}
\safemath{\veco}{\bmo}
\safemath{\vecp}{\bmp}
\safemath{\vecq}{\bmq}
\safemath{\vecr}{\bmr}
\safemath{\vecs}{\bms}
\safemath{\vect}{\bmt}
\safemath{\vecu}{\bmu}
\safemath{\vecv}{\bmv}
\safemath{\vecw}{\bmw}
\safemath{\vecx}{\bmx}
\safemath{\vecy}{\bmy}
\safemath{\vecz}{\bmz}

\safemath{\veczero}{\bmzero}
\safemath{\vecone}{\bmone}
\safemath{\vecxi}{\bmxi}
\safemath{\veclambda}{\bmlambda}
\safemath{\vecmu}{\bmmu}
\safemath{\vectheta}{\bmtheta}
\safemath{\vecphi}{\bmphi}
\safemath{\vecdelta}{\bmdelta}

\safemath{\matA}{\bA}
\safemath{\matB}{\bB}
\safemath{\matC}{\bC}
\safemath{\matD}{\bD}
\safemath{\matE}{\bE}
\safemath{\matF}{\bF}
\safemath{\matG}{\bG}
\safemath{\matH}{\bH}
\safemath{\matI}{\bI}
\safemath{\matJ}{\bJ}
\safemath{\matK}{\bK}
\safemath{\matL}{\bL}
\safemath{\matM}{\bM}
\safemath{\matN}{\bN}
\safemath{\matO}{\bO}
\safemath{\matP}{\bP}
\safemath{\matQ}{\bQ}
\safemath{\matR}{\bR}
\safemath{\matS}{\bS}
\safemath{\matT}{\bT}
\safemath{\matU}{\bU}
\safemath{\matV}{\bV}
\safemath{\matW}{\bW}
\safemath{\matX}{\bX}
\safemath{\matY}{\bY}
\safemath{\matZ}{\bZ}
\safemath{\matzero}{\bmzero}

\safemath{\matDelta}{\bDelta}
\safemath{\matLambda}{\bLambda}
\safemath{\matPhi}{\bPhi}
\safemath{\matSigma}{\bSigma}
\safemath{\matOmega}{\bOmega}
\safemath{\matTheta}{\bTheta}

\safemath{\matidentity}{\matI}
\safemath{\matone}{\matO}

\safemath{\rnda}{A}
\safemath{\rndb}{B}
\safemath{\rndc}{C}
\safemath{\rndd}{D}
\safemath{\rnde}{E}
\safemath{\rndf}{F}
\safemath{\rndg}{G}
\safemath{\rndh}{H}
\safemath{\rndi}{I}
\safemath{\rndj}{J}
\safemath{\rndk}{K}
\safemath{\rndl}{L}
\safemath{\rndm}{M}
\safemath{\rndn}{N}
\safemath{\rndo}{O}
\safemath{\rndp}{P}
\safemath{\rndq}{Q}
\safemath{\rndr}{R}
\safemath{\rnds}{S}
\safemath{\rndt}{T}
\safemath{\rndu}{U}
\safemath{\rndv}{V}
\safemath{\rndw}{W}
\safemath{\rndx}{X}
\safemath{\rndy}{Y}
\safemath{\rndz}{Z}

\safemath{\rveca}{\bimA}
\safemath{\rvecb}{\bimB}
\safemath{\rvecc}{\bimC}
\safemath{\rvecd}{\bimD}
\safemath{\rvece}{\bimE}
\safemath{\rvecf}{\bimF}
\safemath{\rvecg}{\bimG}
\safemath{\rvech}{\bimH}
\safemath{\rveci}{\bimI}
\safemath{\rvecj}{\bimJ}
\safemath{\rveck}{\bimK}
\safemath{\rvecl}{\bimL}
\safemath{\rvecm}{\bimM}
\safemath{\rvecn}{\bimN}
\safemath{\rveco}{\bomO}
\safemath{\rvecp}{\bimP}
\safemath{\rvecq}{\bimQ}
\safemath{\rvecr}{\bimR}
\safemath{\rvecs}{\bimS}
\safemath{\rvect}{\bimT}
\safemath{\rvecu}{\bimU}
\safemath{\rvecv}{\bimV}
\safemath{\rvecw}{\bimW}
\safemath{\rvecx}{\bimX}
\safemath{\rvecy}{\bimY}
\safemath{\rvecz}{\bimZ}

\safemath{\rvecxi}{\bmxi}
\safemath{\rveclambda}{\bmlambda}
\safemath{\rvecmu}{\bmmu}
\safemath{\rvectheta}{\bmtheta}
\safemath{\rvecphi}{\bmphi}

\safemath{\rmatA}{\bimA}
\safemath{\rmatB}{\bimB}
\safemath{\rmatC}{\bimC}
\safemath{\rmatD}{\bimD}
\safemath{\rmatE}{\bimE}
\safemath{\rmatF}{\bimF}
\safemath{\rmatG}{\bimG}
\safemath{\rmatH}{\bimH}
\safemath{\rmatI}{\bimI}
\safemath{\rmatJ}{\bimJ}
\safemath{\rmatK}{\bimK}
\safemath{\rmatL}{\bimL}
\safemath{\rmatM}{\bimM}
\safemath{\rmatN}{\bimN}
\safemath{\rmatO}{\bimO}
\safemath{\rmatP}{\bimP}
\safemath{\rmatQ}{\bimQ}
\safemath{\rmatR}{\bimR}
\safemath{\rmatS}{\bimS}
\safemath{\rmatT}{\bimT}
\safemath{\rmatU}{\bimU}
\safemath{\rmatV}{\bimV}
\safemath{\rmatW}{\bimW}
\safemath{\rmatX}{\bimX}
\safemath{\rmatY}{\bimY}
\safemath{\rmatZ}{\bimZ}

\safemath{\rmatDelta}{\bimDelta}
\safemath{\rmatLambda}{\bimLambda}
\safemath{\rmatPhi}{\bimPhi}
\safemath{\rmatSigma}{\bimSigma}
\safemath{\rmatOmega}{\bimOmega}
\safemath{\rmatTheta}{\bimTheta}

\usepackage{amssymb}
\usepackage{amsfonts}
\usepackage{mathrsfs}
\usepackage{xspace}
\usepackage{bm}
\usepackage{fancyref}
\usepackage{textcomp}

\usepackage{multirow}
\usepackage{stmaryrd}

\newenvironment{textbmatrix}{	\setlength{\arraycolsep}{2.5pt}%
								\big[\begin{matrix}}{\end{matrix}\big]%
								\raisebox{0.08ex}{\vphantom{M}}}

\def\be{\begin{equation}}
\def\ee{\end{equation}}
\def\een{\nonumber \end{equation}}
\def\mat{\begin{bmatrix}}
\def\emat{\end{bmatrix}}
\def\btm{\begin{textbmatrix}}
\def\etm{\end{textbmatrix}}

\def\ba#1\ea{\begin{align}#1\end{align}}
\def\bas#1\eas{\begin{align*}#1\end{align*}}
\def\bs#1\es{\begin{split}#1\end{split}}
\def\bg#1\eg{\begin{gather}#1\end{gather}}
\def\bml#1\eml{\begin{multline}#1\end{multline}}
\def\bi#1\ei{\begin{itemize}#1\end{itemize}}

\DeclareMathOperator*{\argmin}{arg\;min}		

\safemath{\dirac}{\delta}					
\safemath{\krond}{\dirac}					

\safemath{\upto}{\uparrow}
\safemath{\downto}{\downarrow}
\safemath{\iu}{j}							
\safemath{\ev}{\lambda}						
\safemath{\hilseqspace}{l^{2}}				
\newcommand{\banachfunspace}[1]{\setL^{#1}}	
\safemath{\hilfunspace}{\banachfunspace{2}}	

\safemath{\SNR}{\textit{SNR}} 				
\safemath{\PAR}{\textit{PAR}} 				
\safemath{\No}{N_0}							
\safemath{\Es}{E_s}							
\safemath{\Eb}{E_b}							
\safemath{\EbNo}{\frac{\Eb}{\No}}
\safemath{\EsNo}{\frac{\Es}{\No}}

\DeclareMathOperator{\CHop}{\ensuremath{\opH}} 
\safemath{\tvir}{\rndh_{\CHop}}				
\safemath{\tvtf}{\rndl_{\CHop}}				
\safemath{\spf}{\rnds_{\CHop}}				
\safemath{\bff}{H_{\CHop}}					

\safemath{\ircf}{r_{h}}						
\safemath{\tftvcf}{r_{s}}					
\safemath{\tfcf}{r_{l}}						
\safemath{\bfcf}{r_{H}}						

\safemath{\tcorr}{c_h}						
\safemath{\scf}{c_{s}}						
\safemath{\tfcorr}{c_{l}}					
\safemath{\fcorr}{c_{H}}						

\safemath{\mi}{I}							
\safemath{\capacity}{C}						

\safemath{\normal}{\mathcal{N}}			
\safemath{\jpg}{\mathcal{CN}}			
\safemath{\mchain}{\leftrightarrow}		

\safemath{\dB}{\,\mathrm{dB}}
\safemath{\dBm}{\,\mathrm{dBm}}
\safemath{\Hz}{\,\mathrm{Hz}}
\safemath{\kHz}{\,\mathrm{kHz}}
\safemath{\MHz}{\,\mathrm{MHz}}
\safemath{\GHz}{\,\mathrm{GHz}}
\safemath{\s}{\,\mathrm{s}}
\safemath{\ms}{\,\mathrm{ms}}
\safemath{\mus}{\,\mathrm{\text{\textmu}s}}
\safemath{\ns}{\,\mathrm{ns}}
\safemath{\ps}{\,\mathrm{ps}}
\safemath{\meter}{\,\mathrm{m}}
\safemath{\mm}{\,\mathrm{mm}}
\safemath{\cm}{\,\mathrm{cm}}
\safemath{\m}{\,\mathrm{m}}
\safemath{\W}{\,\mathrm{W}}
\safemath{\mW}{\, \mathrm{mW}}
\safemath{\J}{\,\mathrm{J}}
\safemath{\K}{\,\mathrm{K}}
\safemath{\bit}{\,\mathrm{bit}}
\safemath{\nat}{\,\mathrm{nat}}

\safemath{\define}{\triangleq}			

\safemath{\equivalent}{\sim}
\safemath{\distas}{\sim}					
\safemath{\sdiff}{\Delta}				

\safemath{\reals}{\mathbb{R}}
\safemath{\positivereals}{\reals_{+}}
\safemath{\integers}{\mathbb{Z}}
\safemath{\posint}{\integers_{+}}
\safemath{\naturals}{\mathbb{N}}
\safemath{\posnaturals}{\naturals_{+}}
\safemath{\complexset}{\mathbb{C}}
\safemath{\rationals}{\mathbb{Q}}

\newcommand*{\fancyrefapplabelprefix}{app}		
\newcommand*{\fancyrefthmlabelprefix}{thm}		
\newcommand*{\fancyreflemlabelprefix}{lem}		
\newcommand*{\fancyrefcorlabelprefix}{cor}		
\newcommand*{\fancyrefdeflabelprefix}{def}		
\newcommand*{\fancyrefproplabelprefix}{prop}		
\newcommand*{\fancyrefexmpllabelprefix}{exmpl}
\newcommand*{\fancyrefalglabelprefix}{alg}		
\newcommand*{\fancyreftbllabelprefix}{tbl}		

\frefformat{vario}{\fancyrefseclabelprefix}{Section~#1}
\frefformat{vario}{\fancyrefthmlabelprefix}{Theorem~#1}
\frefformat{vario}{\fancyreftbllabelprefix}{Table~#1}
\frefformat{vario}{\fancyreflemlabelprefix}{Lemma~#1}
\frefformat{vario}{\fancyrefcorlabelprefix}{Corollary~#1}
\frefformat{vario}{\fancyrefdeflabelprefix}{Definition~#1}
\frefformat{vario}{\fancyreffiglabelprefix}{Figure~#1}
\frefformat{vario}{\fancyrefapplabelprefix}{Appendix~#1}
\frefformat{vario}{\fancyrefeqlabelprefix}{(#1)}
\frefformat{vario}{\fancyrefproplabelprefix}{Proposition~#1}
\frefformat{vario}{\fancyrefexmpllabelprefix}{Example~#1}
\frefformat{vario}{\fancyrefalglabelprefix}{Algorithm~#1}

\safemath{\dictab}{[\,\dicta\,\,\dictb\,]}

\safemath{\ysig}{\bmy}
\safemath{\ysighat}{\hat{\ysig}}
\safemath{\ysigdim}{M}
\safemath{\xsig}{\bmx}
\safemath{\xsigdim}{N}
\safemath{\nx}{n_x}
\safemath{\zsig}{\bmz}
\safemath{\zsigdim}{\ysigdim}
\safemath{\rsig}{\bmr}
\safemath{\Adict}{\bA}
\safemath{\Adicttilde}{\widetilde{\Adict}}
\safemath{\Adictdim}{\outputdim\times\xsigdim}
\safemath{\avec}{\bma}
\safemath{\avectilde}{\tilde{\avec}}
\safemath{\Bdict}{\bB}
\safemath{\Bdicttilde}{\widetilde{\Bdict}}
\safemath{\Cdict}{\bC}
\safemath{\cvec}{\bmc}
\safemath{\Ddict}{\bD}
\safemath{\Ddictdim}{\ysigdim\times\xsigdim}
\safemath{\dvec}{\bmd}
\safemath{\Ddicttilde}{\widetilde{\bD}}
\safemath{\Bonb}{\bB}
\safemath{\bvec}{\bmb}
\safemath{\Bonbdim}{\ysigdim\times\ysigdim}
\safemath{\noise}{\bmn}
\safemath{\noisedim}{\ysigim}
\safemath{\err}{\bme}
\safemath{\errdim}{\ysigdim}
\safemath{\errset}{\setE}
\safemath{\nerr}{n_e}
\safemath{\delop}{\bP_\errset}
\safemath{\delopc}{\bP_{{\errset}^c}}

\safemath{\cplxi}{\imath}
\safemath{\cplxj}{\jmath}

\safemath{\dict}{\matD}
\safemath{\inputdim}{N}		
\safemath{\outputdim}{M}		
\safemath{\sparsity}{S}	
\safemath{\inputdimA}{{N_a}}	
\safemath{\inputdimB}{{N_b}}	
\safemath{\elemA}{{n_a}}	
\safemath{\elemB}{{n_b}}	
\safemath{\resA}{\matR_a}	
\safemath{\resB}{\matR_b}	
\safemath{\subD}{\matS} 
\safemath{\subA}{\matS_a} 
\safemath{\subB}{\matS_b} 
\safemath{\dicta}{\matA} 	
\safemath{\dictb}{\matB} 	
\safemath{\hollowS}{H}
\safemath{\hollowA}{H_a}
\safemath{\hollowB}{H_b}
\safemath{\cross}{Z}
\safemath{\coh}{\mu_d}			
\safemath{\coha}{\mu_a}			
\safemath{\cohb}{\mu_b}			
\safemath{\mubs}{\nu}	
\safemath{\cohm}{\mu_m} 
\safemath{\dictset}{\setD}	
\safemath{\dictsetp}{\dictset(\coh,\coha,\cohb)}	
\safemath{\dictsetgen}{\dictset_\text{gen}}
\safemath{\dictsetgenp}{\dictsetgen(\coh)}
\safemath{\dictsetonb}{\dictset_\text{onb}}
\safemath{\dictsetonbp}{\dictsetonb(\coh)}

\safemath{\leftside}{U}
\safemath{\rightsideA}{R_a}
\safemath{\rightsideB}{R_b}

\safemath{\indexS}{\setI_S} 

\safemath{\na}{n_a}			
\safemath{\nb}{n_b}			
\safemath{\coeffa}{p_i}	
\safemath{\coeffb}{q_j}	
\safemath{\seta}{\setP}		
\safemath{\setb}{\setQ}     
\safemath{\setw}{\setW}	
\safemath{\setz}{\setZ}	
\safemath{\cola}{\veca}		
\safemath{\colb}{\vecb}		
\safemath{\cold}{\vecd}		
\safemath{\inputvec}{\vecx} 	
\safemath{\error}{\vece}	
\safemath{\noiseout}{\vecz} 	
\safemath{\inputvecel}{x}
\safemath{\inputveca}{\vecx_a}
\safemath{\inputvecb}{\vecx_b}
\safemath{\outputvec}{\vecy}	
\safemath{\lambdamin}{\lambda_{\mathrm{min}}}

\safemath{\elltwo}{\ell_2}
\safemath{\ellone}{\ell_1}
\safemath{\ellzero}{\ell_0}
\safemath{\ellinf}{\ell_\infty}
\safemath{\ellinftilde}{\ell_{\widetilde\infty}}
\safemath{\licard}{Z(\coh,\coha,\cohb)}
\safemath{\xsol}{\hat{x}}
\safemath{\xbord}{x_b}		
\safemath{\xstat}{x_s}		
\safemath{\xstatLone}{\tilde{x}_s}
\safemath{\order}{\mathcal{O}} 
\safemath{\scales}{\Theta} 
\safemath{\ones}{\mathbf{1}} 
\safemath{\zeroes}{\mathbf{0}} 
\safemath{\thlone}{\kappa(\coh,\cohb)} 
\safemath{\constoneA}{\delta} 
\safemath{\constoneB}{\epsilon} 
\safemath{\nlarge}{L}				   
\safemath{\sumlarge}{S_\nlarge}
\safemath{\maxlarger}{P_\nlarge}	   
\safemath{\Pzero}{\textrm{P0}}	
\safemath{\Pone}{\textrm{P1}}
\safemath{\vecfir}{\vecw}			 
\safemath{\vecsec}{\vecz}
\safemath{\elvecfir}{w}              
\safemath{\elvecsec}{z}				 
\safemath{\nlargefir}{n}
\safemath{\normout}{\gamma}
\safemath{\auxfun}{h}
\safemath{\supp}{\textrm{supp}}

\safemath{\indexa}{\ell}
\safemath{\indexb}{r}
\safemath{\indexc}{i}
\safemath{\indexd}{j}

\safemath{\project}{P}

\allowdisplaybreaks 

\safemath{\LAMA}{\textrm{LAMA}}
\safemath{\MRT}{\textrm{MRT}}
\safemath{\betamax}{\beta^\text{max}_\setO}
\safemath{\betamaxno}{\beta^\text{max}}
\safemath{\betamin}{\beta^\text{min}_\setO}
\safemath{\betaminno}{\beta^\text{min}}

\safemath{\Nomin}{\No^\textnormal{min}(\beta)}
\safemath{\Nominnobeta}{\No^\text{min}}
\safemath{\Nomax}{\No^\textnormal{max}(\beta)}
\safemath{\Nomaxnobeta}{\No^\textnormal{max}}
\safemath{\EX}{E_\textnormal{x}}
\safemath{\EXP}{\EX^\textnormal{p}}
\safemath{\Eo}{E_0}

\safemath{\tmax}{{t_\textnormal{max}}}
\safemath{\MAP}{\textrm{MAP}}
\safemath{\IO}{\textrm{IO}}
\safemath{\JO}{\textrm{JO}}
\safemath{\Nopost}{N_{0}^\textnormal{post}}
\safemath{\MT}{U}
\safemath{\MR}{B}
\safemath{\Tran}{\textnormal{T}}
\safemath{\Herm}{\textnormal{H}}
\safemath{\row}{\textnormal{r}}
\safemath{\col}{\textnormal{c}}

\safemath{\NT}{N_\textnormal{T}}
\safemath{\DSNR}{\delta \textnormal{SNR}}
\safemath{\betaMOR}{\beta^{\star}}

\begin{document}
	
\title{A Deep-Unfolding-Optimized Coordinate-Descent Data-Detector ASIC for mmWave Massive MIMO}
\author{Zixiao Li, Seyed Hadi Mirfarshbafan, Oscar Casta\~neda, and Christoph Studer%
\thanks{Z. Li is with the Department of Information Technology and Electrical Engineering, ETH Zurich, and with the Institute of Neuroinformatics, University of Zurich and ETH Zurich. (email: zixili@ethz.ch)}
\thanks{S.~H.~Mirfarshbafan, O. Casta\~neda, and C.~Studer are with the Department of Information Technology and Electrical Engineering, ETH Zurich. (email: mirfarshbafan@iis.ee.ethz.ch, caoscar@ethz.ch, studer@ethz.ch)}
\thanks{This paper summarizes and extends the results from Z. Li's MSc Thesis; the MSc Thesis and its results are not published elsewhere.}
}

\maketitle

\begin{abstract}

We present a 22\,nm FD-SOI (fully depleted silicon-on-insulator) application-specific integrated circuit (ASIC) implementation of a novel soft-output Gram-domain block coordinate descent (GBCD) data detector for massive multi-user (MU) multiple-input multiple-output (MIMO) systems. The ASIC simultaneously addresses the high throughput requirements for millimeter wave (mmWave) communication, stringent area and power budget per subcarrier in an orthogonal frequency-division multiplexing (OFDM) system, and error-rate performance challenges posed by realistic mmWave channels.  
The proposed GBCD algorithm utilizes a posterior mean estimate~(PME) denoiser and is optimized using deep unfolding, which results in superior error-rate performance even in scenarios with highly correlated channels or where the number of user equipment~(UE) data streams is comparable to the number of basestation~(BS) antennas.
The fabricated GBCD ASIC supports up to 16 UEs transmitting QPSK to 256-QAM symbols to a 128-antenna BS, and achieves a peak throughput of 7.1\,Gbps at 367\,mW. The core area is only 0.97\,mm$^{\bf 2}$ thanks to a reconfigurable array of processing elements that enables extensive resource sharing. Measurement results demonstrate that the proposed GBCD data-detector ASIC achieves best-in-class throughput and area efficiency.

\end{abstract}


\section{Introduction}
\label{sec:intro}

\IEEEPARstart{M}{assive} multiuser (MU) multiple-input multiple-output (MIMO) technology and millimeter-wave (mmWave) communication are critical physical layer technologies in 5th generation (5G) wireless communication systems~\cite{3gpp22, larsson14a, rappaport15a}.
Beyond-5G systems are also believed to build upon these technologies as their combination promises unprecedented  data rates and spectral efficiency. 
In order to deal with the intersymbol interference caused by multipath, orthogonal frequency division multiplexing (OFDM) remains to be the technology of choice, e.g., 
5G new radio~(NR) specifies it as the main waveform~\cite{TS38_211_v17}. 
However, hardware implementations for some of the key baseband processing tasks in mmWave massive MU-MIMO-OFDM systems~\cite{Bolcskei06}, including data detection in the uplink---which is the main focus of this paper---face several critical implementation challenges for the following~reasons:

\begin{itemize}
\item The large number of basestation (BS) antennas and simultaneously transmitting user equipments (UEs) requires processing of high-dimensional data.
\item Wideband mmWave communication requires high baseband sampling rates, which results in a large amount of data to be processed per second.
\item In OFDM systems, data detection and its preprocessing require massively parallel processing over many subcarriers, which necessitates highly efficient hardware accelerators.
\end{itemize}
These challenges require the design of efficient data detectors that keep the costs and power consumption of corresponding BS implementations within reasonable bounds. 

Data detection in multi-antenna wireless systems typically consists of (i) \emph{preprocessing}, which are operations that only depend on channel state information, and (ii) \emph{equalization}, which corresponds to estimating the transmitted symbols using quantities computed in the preprocessing stage. 
{As channel state information is approximately static over an OFDM coherence block consisting of $N_T$ consecutive time slots and $N_S$ consecutive subcarriers, the equalizer processes $T=N_T \times N_S$ receive vectors after a single preprocessing step.}
Since equalization cannot start until preprocessing is completed, the preprocessing latency translates into the need for buffering receive vectors before they can be fed into the equalizer~\cite{Burg06}. 
Furthermore, since preprocessing must be performed only when the channel changes, an efficient data-detector design should maximize hardware-resource sharing between preprocessing and equalization. 
In summary, a massive MU-MIMO data detector that addresses the above-mentioned implementation challenges should have the following properties: (i) small area and low power, such that parallel independent data-detector cores (each responsible for one or several subcarriers) can be instantiated without resulting in excessive silicon area or power; (ii)  sufficiently high throughput in order to complete all data-detection tasks within the required time constraint; and (iii) low preprocessing latency to minimize data buffering. Needless to mention, the data detection algorithm should achieve excellent error-rate performance, even for ill-conditioned channel realizations that are common at mmWave frequencies~\cite{rappaport15a}.

In what follows, we adopt an algorithm-hardware co-design approach to design a massive MU-MIMO data-detector core that meets the above listed criteria. Specifically, we propose a novel data detection algorithm that enables the design of a very large-scale integration (VLSI) implementation that achieves high area efficiency at competitive energy efficiency, so that one data-detector core can then be replicated to process the data of all used subcarriers in parallel.

\subsection{Prior Work}
To enable high throughput data detection for massive MU-MIMO systems, a number of ASIC implementations have been proposed~\cite{prabhu17, jeon19b, tang18, tang19, peng18b, liu20}. Unfortunately, some of these designs do not scale well to OFDM systems with a large number of subcarriers due to either a relatively large implementation area per data-detection core~\cite{tang18, peng18b, liu20}, low energy efficiency~\cite{jeon19b, peng18b}, or limited throughput~\cite{prabhu17, jeon19b}. Another strain of designs has achieved high area efficiency by using low-complexity algorithms that rely on the channel-hardening assumption~\cite{tang19, liu20}. This assumption describes the phenomenon in massive MU-MIMO systems by which the correlation between UE channels reduces when increasing the number  of BS antennas~$B$ for a fixed number  of UEs~$U$~\cite{Willhammar18}. 
However, channel hardening does \emph{not} hold for smaller $B/U$ ratios, nor in channels with strong correlation or with  strong line-of-sight (LoS) components between the UEs and the BS.
The latter case is typical at mmWave frequencies as wave propagation is predominantly directional with few scattered paths arriving at the receiver. Therefore, data detectors that rely on channel hardening typically perform poorly under realistic mmWave channels.

For many existing data detectors, the preprocessing stage involves computation of the Gram matrix $\bG = \bH^\Herm \bH$, where $\bH \in \mathbb{C}^{B \times U}$ is the estimated channel matrix at the BS.
Zero-forcing (ZF) and linear minimum mean squared error (LMMSE) equalization are two prominent data detection algorithms in massive MU-MIMO as they require low complexity and achieve acceptable error-rate performance.
Such algorithms, however, require implicit or explicit inversion of the $U \times U$ (regularized) Gram matrix and thereby, incur large area and high preprocessing latency~\cite{wu17a, prabhu17}.
To reduce preprocessing complexity, many low-complexity iterative algorithms have been proposed; see, e.g.,~\cite{gao14, wu2014large, Zhu15, wu17, dai15}.
A promising low-complexity data detector based on coordinate descent (CD), referred to as optimized coordinate descent (OCD), has been proposed in~\cite{wu16}. The main advantage of this method is its low preprocessing complexity as no Gram matrix must be computed, which results in low  latency. However, as we will show in \fref{sec:performance}, the performance of OCD severely degrades under LoS channels. Moreover, our complexity analysis in \fref{sec:complexity} reveals that the total complexity of OCD, which includes both preprocessing and equalization, increases rapidly with the number of transmissions~$T$ per coherence block due to its relatively high equalization complexity.

Recently, there has been extensive research on the application of deep unfolding to optimize the performance of MU-MIMO data-detection algorithms~\cite{hershey2014deep, stimming19}. This approach unfolds iterative algorithms into a neural-network-like structure and employs deep-learning techniques to train certain algorithm hyperparameters. For example, DetNet~\cite{Samuel19} unfolded projected gradient descent and OAMPNet~\cite{He20} optimized orthogonal approximate message passing with only four parameters to be learned per layer. MMNet~\cite{Khani20} further reduced the complexity of OAMPNet by replacing the matrix inverse operation with a trainable matrix. Furthermore, HyperEPNet~\cite{zhang22epnet} utilized a hypernetwork to train the damping factors of an expectation-propagation algorithm. Despite the potential performance advantages of deep unfolding, such existing deep-unfolding-based detectors have exhibited prohibitive complexity for hardware implementation. DetNet requires a large network with 1\,M to 10\,M parameters, while the complexity of OAMPNet and HyperEPNet for one iteration scales with $O(B^3)$ and $O(U^3)$, respectively, due to the necessary matrix inversion in the unfolded algorithm. Although MMNet avoided a matrix inversion step, its per-iteration complexity scales with $O(B^2)$, which is unsuitable for massive MIMO systems with a large number of BS antennas. In addition, MMNet's online learning strategy is impractical for hardware because of its high resource requirements. Therefore, the development of algorithms that harness the advantages of deep unfolding while leading to hardware-friendly architectures remains to be relevant.

\subsection{Contributions}
Motivated by the low preprocessing complexity of OCD and the advantages of deep unfolding, we propose the Gram-domain block coordinate descent algorithm (GBCD) with a deep-unfolding-assisted PME denoiser, which overcomes the key limitations of OCD. Additionally, we present a corresponding area- and energy-efficient VLSI architecture and measurement results for a fabricated application-specific integrated circuit (ASIC) in a $22$\,nm fully-depleted silicon-on-insulator (FD-SOI) process.
The resilience of GBCD against ill-conditioned channels, as well as its low area and low power, render it particularly attractive for massive MU-MIMO-OFDM systems. The key contributions of the paper are summarized as follows: 
\begin{itemize}
\item \emph{Algorithm design:} We propose the GBCD soft-output data detector by combining block coordinate descent (BCD)~\cite{Seidel19} with a Gram-domain transformation to significantly reduce the equalization complexity and the latency of CD. BCD is further combined with a UE-sorting scheme, resulting  in robust detection performance, even under realistic channels with strong correlation. We also utilize an effective posterior mean estimate (PME)-based denoiser with a hardware-friendly piecewise linear approximation, and we optimize all of our algorithm's parameters via deep unfolding.
\item \emph{VLSI architecture:} We design a VLSI architecture for the soft-output GBCD data detector, which supports both BOX and PME denoisers, and constellations ranging from QPSK to 256-QAM. The VLSI architecture achieves high area efficiency by reusing processing elements for preprocessing and equalization. 
\item \emph{ASIC measurement:} We provide measurement results for a fabricated ASIC, and we demonstrate that our data detector achieves best-in-class area efficiency among the existing ASICs that are able to achieve good error-rate performance under mmWave channel conditions. 

\end{itemize}

\subsection{Notation}
Boldface lowercase and uppercase letters represent column vectors and matrices, respectively; sets are written as calligraphic letters. 
For a matrix $\bG$, $\bG_\setA$ is the submatrix formed by the columns of $\bG$ indexed by the set $\setA$, and $\bG_{\setA,\setB}$ is the submatrix formed by the elements of $\bG$ that are in the rows indexed by  $\setA$ and the columns indexed by $\setB$.
The Hermitian transpose of the matrix $\bG$ is denoted by $\bG^\Herm$ and the entry in the $m$th row and $n$th column  is $G_{m,n}$.
For a vector $\bmg$, the $k$th entry is denoted by $g_k = [\bmg]_k$,  and its Euclidean norm by~$\|\vecg\|$. The real and imaginary parts of a complex number~$z\in\complexset$ are $z^\textnormal{R}$ and $z^\textnormal{I}$, respectively.
The $M\times N$ all-zeros and $N\times N$ identity matrix are $\bZero_{M\times N}$ and~$\bI_N$, respectively.
$O(\cdot)$ denotes the big-O complexity order and $\mathbb{E}[\cdot]$ denotes expectation.

\subsection{Paper Outline}
The rest of the paper is organized as follows. \fref{sec:algo} proposes the GBCD algorithm with the PME denoiser and provides a complexity and performance analysis. \fref{sec:vlsi} details our VLSI architecture and \fref{sec:asic} presents ASIC measurement results along with a comparison of state-of-the-art data detectors. \fref{sec:conclusion} concludes the paper.

\section{\textsc{GBCD: Gram Block Coordinate Descent}}
\label{sec:algo} 

\subsection{System Model} \label{sec:sysmodel}

We consider the uplink of a massive MU-MIMO system in which $U$ single-antenna UEs\footnote{In the case of multi-antenna UEs, each transmitting one or multiple data streams, the system model can be transformed into the form of \fref{eq:MIMOsys}, in which the channel matrix $\bH$ absorbs the effect of beamforming at each UE.} concurrently transmit OFDM-modulated signals to a $B$-antenna BS. Assuming perfect synchronization and a sufficiently long cyclic prefix, OFDM transmission decomposes the wireless channel into $W$ parallel and independent subcarriers in the frequency domain. As each data-detection problem for every subcarrier is independent, we will focus on a single OFDM subcarrier to simplify notation. 
Each UE's data stream is independently encoded at a rate~$R$ and mapped to a constellation $\mathcal{O}$ (e.g., $256$-QAM) of cardinality~$Q$. The transmit vector $\bms\in \mathcal{O}^{U}$ contains the symbols transmitted by all UEs, which we assume to be i.i.d.\ zero mean with variance~\Es, so that $\mathbb{E}_{\bms}[\bms \bms^\Herm]=\Es \bI_U$. The input-output relation for a single subcarrier is given by
\begin{align} \label{eq:MIMOsys}
    \bmy = \bH\bms + \bmn,
\end{align}
where $\bmy\in \mathbb{C}^B$ is the receive signal vector, $\bH\in \mathbb{C}^{B\times U}$ is the uplink MIMO channel matrix, and $\bmn\in \mathbb{C}^B$ models i.i.d. circularly-symmetric complex Gaussian noise with variance~\No. 

\subsection{GBCD Detector}
\label{sec:GBCDdetector}
Optimal MIMO data detection in terms of minimizing the vector error rate is achieved by the maximum-likelihood (ML) data detector. For discrete constellations, ML data detection corresponds to an optimization problem with discrete constraints, which results in an NP-hard problem that cannot be solved efficiently for large~$U$.
As a remedy, a plethora of approximate algorithms have been proposed to relax the discrete constraint of the ML optimization problem to a set of convex constraints. One promising candidate is BOX equalization, which solves the following optimization problem~\cite{wu16}:
\begin{align} \label{eq:BOX}
	\hat{\bms}^{\text{BOX}}=\argmin_{\bmz \in \mathcal{C}_{\mathcal{O}}^U}{\| \bmy - \bH \bmz\|^2}.
\end{align}
Here, $\mathcal{C}_{\mathcal{O}}$ is the tightest convex polytope around the scalar constellation set $\mathcal{O}$. To solve~\fref{eq:BOX}, the work in~\cite{wu16} proposes to apply CD~\cite{Wright15}, resulting in the OCD data-detection algorithm. Although the original OCD algorithm significantly reduces preprocessing complexity and requires low area in its implementation, it has three key limitations: (i) high equalization complexity per transmission, (ii) increased latency due to sequential processing, and (iii) degraded error-rate performance under correlated channels or for smaller $B/U$ ratios. While~\cite{Seidel19} introduces BCD to decrease the detection latency of CD, the other two challenges persist. 
To address all of these challenges, we propose the GBCD detector to solve \fref{eq:BOX} by combining BCD with a Gram-domain transformation and SINR-based UE sorting. We further improve our algorithm's performance by using a deep-unfolding-assisted PME denoiser, as discussed in \fref{sec:PME}.

\subsubsection{BCD with the BOX denoiser}
The BCD algorithm with the BOX denoiser is the starting point of our proposed algorithm.
To derive BCD, we divide the UEs into $M$ blocks (or groups) $\{ \mathcal{A}_{1}, \mathcal{A}_{2},\ldots, \mathcal{A}_{M} \}$ of size $L=|\mathcal{A}_{m}|=\frac{U}{M}, m=1,2,\ldots,M$. 
BCD tries to find the optimal solution $\hat{\bmz}_{\mathcal{A}_{m}}$ for the $m$th block $\mathcal{A}_{m}$ at a time in~\fref{eq:BOX}, while all other blocks are fixed, which is equivalent to solving the following optimization problem: 
\begin{equation} \label{eq:BOX_BLOCK}
\hat{\bmz}_{\mathcal{A}_{m}} = \argmin_{\bmz_{\mathcal{A}_{m}} \in \mathcal{C}_{\mathcal{O}}^L}{\| \bmy - \bH \bmz\|^2}.
\end{equation}
\noindent Ignoring the constraint, the least squares solution to \fref{eq:BOX_BLOCK} is
\begin{equation} \label{eq:raw_estimate}
	\bmv_{\mathcal{A}_{m}} = \biggl(\bH_{\mathcal{A}_{m}}^\Herm \bH_{\mathcal{A}_{m}}\biggr)^{\!-1}\bH_{\mathcal{A}_{m}}^\Herm\biggl(\bmy-\!\!\!\sum_{j=1, j\neq m}^{M}\bH_{\mathcal{A}_{j}}\bmz_{\mathcal{A}_{j}}\biggr),
\end{equation}

\noindent where $\bH_{\mathcal{A}_{m}}$ is the matrix containing the columns of $\bH$ indexed by $\mathcal{A}_{m}$. Since the feasible region of~\fref{eq:BOX_BLOCK} is a box around the constellation points, the derivative does not exist at the boundary, so standard gradient-based optimization methods are not applicable. Instead, we use a subgradient-based approach~\cite{wu16} to solve~\fref{eq:BOX_BLOCK} for $\hat{\bmz}_{\mathcal{A}_{m}}$ as
\begin{equation} \label{eq:BCD_updaterule}
\hat{\bmz}_{\mathcal{A}_{m}} = \mathcal{B}_{\mathcal{O}}(\bmv_{\mathcal{A}_{m}}),
\end{equation} 
where $\mathcal{B}_{\mathcal{O}}(\cdot)$ is the element-wise BOX denoiser. For a complex-valued scalar input $v$, this denoiser computes
\begin{align} \label{eq:BOXdenoiser}
\mathcal{B}_{\mathcal{O}}(v) = \begin{cases} v & \text{if} ~ v  \in \mathcal{C}_{\mathcal{O}}, \\
\argmin_{q \in \mathcal{C}_{\mathcal{O}}} |v - q| & \text{if} ~  v  \notin \mathcal{C}_{\mathcal{O}}. \end{cases}
\end{align}

\subsubsection{Gram-domain BCD (GBCD)}
GBCD performs the same operations as detailed above for BCD, with the difference that it factors out the Gram matrix $\bG=\bH^\Herm \bH$ and the matched filter (MF) vector $\bmy^{\text{MF}}=\bH^\Herm \bmy$, which  are utilized in all iterations of BCD as revealed by reformulating \fref{eq:raw_estimate} to
\begin{align} 
\bmv_{\mathcal{A}_{m}} &= \left(\bH_{\mathcal{A}_{m}}^\Herm \bH_{\mathcal{A}_{m}}\right)^{\!-1}\bH_{\mathcal{A}_{m}}^\Herm \biggl(\bmy- \bH \bmz +\bH_{\mathcal{A}_{m}}\bmz_{\mathcal{A}_{m}}\biggr) \\
&= \left(\bG_{\mathcal{A}_m,\mathcal{A}_m}\right)^{\!-1} \!\!\biggl(\bmy^{\text{MF}}_{\mathcal{A}_{m}}\!\!-\!\bH_{\mathcal{A}_{m}}^\Herm \bH\bmz\!+\!\bG_{\mathcal{A}_m,\mathcal{A}_m}\bmz_{\mathcal{A}_{m}}\!\!\biggr) \\
&=  \bK_m \bmr_{\mathcal{A}_{m}}+\bmz_{\mathcal{A}_{m}}. \label{eq:vfinal}
\end{align}
Here, $\bmy^{\text{MF}}_{\mathcal{A}_{m}}$ denotes the entries of $\bmy^{\text{MF}}$ indexed by $\mathcal{A}_{m}$, $\bG_{\mathcal{A}_m,\mathcal{A}_m}$ is the submatrix of $\bG$ with rows and columns indexed by $\mathcal{A}_{m}$, and $\bmr_{\mathcal{A}_{m}}=\bmy^{\text{MF}}_{\mathcal{A}_{m}}\!\!-\!\bH_{\mathcal{A}_{m}}^\Herm \bH\bmz$. With the reformulation in \fref{eq:vfinal}, $\bG$ and the $L \times L$ matrices $\bK_m=(\bG_{\mathcal{A}_m,\mathcal{A}_m})^{-1}$ are precomputed during preprocessing, and $\bmr=\bmy^{\text{MF}}-\bG\bmz$ is defined as the residual vector that keeps track of the remaining noise plus interference terms throughout the GBCD iterations.

One outer iteration of GBCD consists of sequentially applying \fref{eq:vfinal} and  \fref{eq:BCD_updaterule} for all blocks $\mathcal{A}_{m}$, $m=1,2,\ldots,M$, where each new estimate $\hat{\bmz}_{\mathcal{A}_{m}}$ is immediately used in computing the estimates of the subsequent blocks. This process is repeated for $K$ outer iterations to obtain the final estimates. 
Concretely, during the $k$th outer iteration, the new estimate for the $m$th block is $\bmz_{\mathcal{A}_{m}}^{(k)} = \mathcal{B}_{\mathcal{O}}(\bmv_{\mathcal{A}_{m}}^{(k)})$, where $\bmv_{\mathcal{A}_{m}}^{(k)}=\bK_m \bmr_{\mathcal{A}_{m}}+\bmz_{\mathcal{A}_{m}}^{(k-1)}$, with $\bmz_{\mathcal{A}_{m}}^{(0)}=0$. Subsequently,
the residual vector $\bmr$ is updated as
\begin{equation}
\bmr\leftarrow \bmr - \bG_{\mathcal{A}_{m}}\Delta \bmz_{\mathcal{A}_{m}}^{(k)},
\end{equation}
where $\Delta \bmz_{\mathcal{A}_{m}}^{(k)} = \bmz_{\mathcal{A}_{m}}^{(k)} - \bmz_{\mathcal{A}_{m}}^{(k-1)}$; $\bmr$ is initialized to $\bmy^{\text{MF}}$.

Both BCD (i.e., simultaneously updating the estimates for a block of UEs) and the Gram-domain transformation (i.e., factoring out $\bG$ and $\bmy^{\text{MF}}$ from the BCD updates in \fref{eq:vfinal}, hence avoiding their computation in every iteration) improve the effectiveness of OCD. The Gram-domain transformation reduces equalization complexity from $O(BU)$ to $O(U^2)$ without affecting performance. BCD improves the error-rate performance, as it applies zero-forcing in \fref{eq:raw_estimate} to mitigate inter-block interference (cf.~\fref{sec:individual_effects}). In the fabricated chip, we set the block size $L=2$, which simplifies the inversion of the matrix~$\bG_{\mathcal{A}_m,\mathcal{A}_m}$. Furthermore, we have observed that larger block sizes do not result in significantly better performance. As we will demonstrate in \fref{sec:complexity} and \fref{sec:performance}, BCD with $L=2$ strikes a good balance between performance and complexity. Furthermore, BCD with $L=2$ roughly halves the equalization latency compared to OCD, as it decreases the number of blocks (or, equivalently, the number of inner iterations) from $U$ to~$U/2$.

\subsubsection{SINR-Based UE Sorting} \label{sec:UESort}

The performance degradation of CD under correlated channels is partly due to error propagation, which results from erroneous estimation results in early iterations. 
Hence, the order of coordinate descent updates is critical for achieving good performance, especially in correlated channels. 
Intuitively, in a successive interference cancellation scheme, we want to first estimate the signals of the UE block with the highest post-equalization signal-to-interference-plus-noise ratio (SINR), subtract the estimated symbols from the residual vector, and proceed to the next UE block with the highest SINR. 
The optimal strategy is to list all possible UE block partitions in each outer iteration, compute the post-equalization SINR of each block, and sort them accordingly. Such an approach would lead to high complexity and long delay. Therefore, we resort to an approximate single-UE SINR metric computed for the first iteration (without any denoising applied) and keep the same ordering in all subsequent iterations. In practice, replacing the block-UE SINR computation with the single-UE SINR approximation incurs no notable performance degradation.
Moreover, direct computation of the SINR typically  involves   divisions by small numbers in case of low noise-plus-interference (NPI) variance; this complicates its calculation with fixed-point arithmetic. 
To improve numerical stability, we rather compute the per-UE reciprocal value of the SINR as
\begin{equation}
\textit{SINR}_u^{-1} = \frac{\lambda_u}{|G_{u,u}|^2}+\frac{\No}{\Es |G_{u,u}|},
\end{equation}
where $\lambda_u=\sum_{i=1, i\neq u}^{U}|G_{u,i}|^2$. Since the values of ~$G_{u,u}$, \mbox{$u=1,2,\ldots,U$}, are positive and generally large, efficient computation of their reciprocal values in hardware can be achieved by using small look-up tables (LUTs).

We sort the UEs in ascending order based on the computed $\textit{SINR}^{-1}_u$, $u=1,2,\ldots,U$, and generate the corresponding index vector $\bm{\nu}=[\nu_1,\ldots,\nu_U]$. This SINR-sorted UE list $\bm{\nu}$ is then used to form UE blocks as \mbox{$\mathcal{A}_m = \{\nu_{(m-1)L+1},\ldots,\nu_{mL}\}$}, \mbox{$m=1,2,\ldots,M$}. For example, the $L$ UEs with the highest SINR are grouped together in the first block $\mathcal{A}_1$ and estimated first. Note that both the SINR values and the UE order remain constant over a coherence block. Therefore, they are computed only once during the preprocessing phase.

\subsubsection{LLR Approximation}
\label{sec:LLR}
Due to the nonlinear denoising operation in GBCD, there is no closed-form expression for the exact computation of the LLR values.
We therefore resort to an approximation based on the max-log LLR values~\cite{studer2011asic} computed for LMMSE equalization
\begin{equation} \label{eq:LLR}
\textit{LLR}_{u,b}=\frac{1}{\xi_{u}}\!\left(\min_{a\in \mathcal{O}_{b}^0}|\hat{s}_{u}-\mu_u a|^2\!-\!\min_{a\in \mathcal{O}_{b}^1}|\hat{s}_{u}-\mu_u a|^2\right)\!,
\end{equation}
where $\hat{s}_{u}$ is the soft estimate of the $u$th UE's symbol, and the sets $\mathcal{O}_{b}^0$ and $\mathcal{O}_{b}^1$ contain the constellation symbols whose $b$th bit is $0$ and $1$, respectively. In GBCD, we set $\hat{\bms}$ to the unconstrained estimate $\bmv^{(K)}$ of the last iteration rather than the denoised estimate $\bmz^{(K)}$, as the denoiser pulls the estimates closer to the constellation points, resulting in overconfident LLR values and poor error-correction performance. 
For the LMMSE detector, the channel gain $\mu_u$ and NPI variance $\xi_u$ in~\fref{eq:LLR} can be computed in closed form as $\mu^{\text{LMMSE}}_u=[{\bA}^{-1}\bG]_{u,u}$  and $\xi^{\text{LMMSE}}_u = E_s (1-\mu_u^{\text{LMMSE}})\mu_u^{\text{LMMSE}}$, where $\bA=\bG+N_0E_s^{-1}{\bI}_U$. These expressions, however, do not apply to GBCD.
In OCD~\cite{wu16}, the channel gains were approximated using the truncated Neumann series expansion from~\cite{yin14b}
\begin{equation} \label{eq:LLRmu}
\mu_u \approx \frac{G_{u,u}}{G_{u,u}+\alpha},
\end{equation}
where $\alpha=N_0E_s^{-1}$ is a normalization factor. For GBCD with the BOX denoiser, we adopt the same approach. However, in \fref{sec:DUscheme}, we will introduce a new approach to adapt the value of  $\alpha$ in order to obtain a more powerful denoiser. We then use the approximation $\xi_u \approx E_s(1-\mu_u)\mu_u$. As we will show in \fref{sec:performance} and \fref{sec:PLM}, this approximation enables both good error-rate performance and efficient computation.

\subsection{Deep-Unfolding-Assisted Posterior Mean Estimate (PME)} \label{sec:PME}
Although the BOX denoiser defined in \fref{eq:BOXdenoiser} is shown to be effective in denoising the unconstrained estimates and is simple to implement in hardware, it ignores the discrete nature of the transmit constellation and only affects estimates that fall outside the convex polytope $\mathcal{C}_{O}$.
To address this issue, we propose to apply an entry-wise PME denoiser~\cite{jeon21, song21, sun24}. In what follows, we focus on the estimate of a single UE in a given iteration, so we omit super- and subscripts.

\subsubsection{PME Function}
\label{sec:pme_exact}
The relationship between the unconstrained symbol estimate $v\in\complexset$ of each UE defined in \fref{eq:raw_estimate} and the true transmit symbol $s\in \mathcal{O}$ can be modeled as
\begin{align}\label{eq:PMElinear}
	v=\beta s+e,   
\end{align}
where $\beta\in\reals$ models the channel gain and $e$ models noise plus interference. 
If $\beta$ and the distribution of $e$ are given, then we can optimally denoise $v$ using the posterior mean estimator  $z = \mathbb{E}\left[ s|v\right]$, which minimizes the mean squared error (MSE) between $s$ and $z$.
Here, we assume that the prior distribution of $s$ is a uniform distribution over the discrete set $\mathcal{O}$, with independent real and imaginary components. We further assume that $e$ is circularly symmetric complex Gaussian, independent of $s$ with precision $\omega/2$ (i.e., variance $2/\omega$). Therefore, we can decompose the complex-valued PME into real and imaginary  components as follows:
\begin{align} \label{eq:exactposmean}
	\mathbb{E}\left[ s|v\right] = \mathbb{E}\left[ s^\textnormal{R}|v^\textnormal{R}\right] + j\mathbb{E}\left[ s^\textnormal{I}|v^\textnormal{I}\right].  
\end{align} 

The real-valued posterior mean function $\mathbb{E}\left[ s^\textnormal{R}|v^\textnormal{R}\right]$ in \fref{eq:exactposmean} corresponds to
\begin{align}
	\mathcal{P}_{\mathcal{O}'}(v^\textnormal{R};\omega, \beta) = \frac{\sum\sb{a\in \mathcal{O}'}a\exp(-\omega(v^\textnormal{R}-\beta a)^2)}{\sum\sb{a\in \mathcal{O}'}\exp(-\omega(v^\textnormal{R}-\beta a)^2)},
	\label{eq:explicitPME}
\end{align}
where $\mathcal{O}'$ is the $\sqrt{Q}$-PAM constellation corresponding to $\mathcal{O}$. The function $\mathcal{P}_{\mathcal{O}}(\cdot)$ operates element-wise on the entries of its input block, and for a complex scalar input~$v$, it outputs $\mathcal{P}_{\mathcal{O}}(v) = \mathcal{P}_{\mathcal{O}'}(v^\textnormal{R}) + j\mathcal{P}_{\mathcal{O}'}(v^\textnormal{I})$.

\subsubsection{Piecewise Linear Approximation for PME}
\label{sec:PMEapprox}

\begin{figure}[t]
	\centering
	\includegraphics[width=0.8\columnwidth]{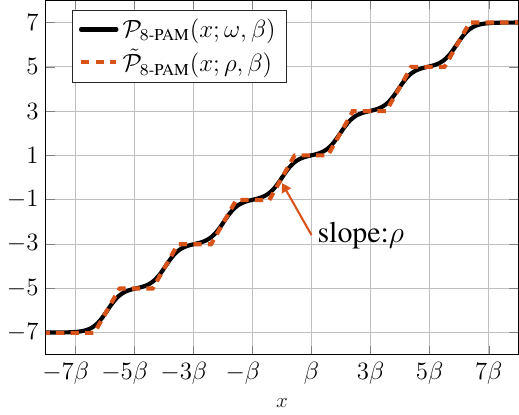}
	\caption{Exact PME $\mathcal{P}_{\mathcal{O}'}(x;\omega, \beta)$ and the proposed piecewise linear approximation $\tilde{\mathcal{P}}_{\mathcal{O}'}(x;\rho, \beta)$ for $8$-PAM with zero-mean Gaussian noise.}
	\label{fig:PLMforPME}
\end{figure}

The PME function in \fref{eq:explicitPME} involves exponentials, which are difficult to compute in hardware and suffer from numerical instability when fixed-point computations are used. To address this issue, a max-log approximation was proposed in~\cite{jeon21,song21}. Nonetheless, this approximation is complex for high-order modulations (e.g., 256-QAM) and results in high latency when implemented in hardware, which would reduce the throughput due to the iterative nature of our algorithm.

Our approach to this issue is as follows. The exact PME in~\fref{eq:explicitPME} resembles a periodic function with sigmoid-like base functions around each PAM constellation point scaled by $\beta$.
We therefore propose a piecewise linear function to approximate~\fref{eq:explicitPME}. The exact and proposed approximate PME functions are shown in \fref{fig:PLMforPME}. 
In the approximate PME, we replace the sigmoid-like function, whose smoothness is controlled by $\omega$, with a piecewise linear approximation with slope parameter $\rho$, and keep the definition of the scaling factor~$\beta$. The piecewise linear approximate PME function is given by
\begin{equation}\label{eq:PMEapprox}
	\tilde{\mathcal{P}}_{\mathcal{O}'}(v^\textnormal{R}; \rho, \beta) = \sum_{k=-\gamma}^{\gamma}f(\rho(v^\textnormal{R}+2\beta k)),
\end{equation}
where $\gamma=\frac{\sqrt{Q}}{2}-1$ for the $\sqrt{Q}$-PAM constellation and the clipping function is defined as $f(x)=\max\{\min\{x,1\},-1\}$. The corresponding hardware implementation is detailed in \fref{sec:PLM}. Next, we will delve into the estimation of the parameters $\rho$ and $\beta$ in \fref{eq:PMEapprox}.

\subsubsection{Deep unfolding scheme}
\label{sec:DUscheme}

\begin{figure*}[t]
    \centering
    \includegraphics[width=0.8\textwidth]{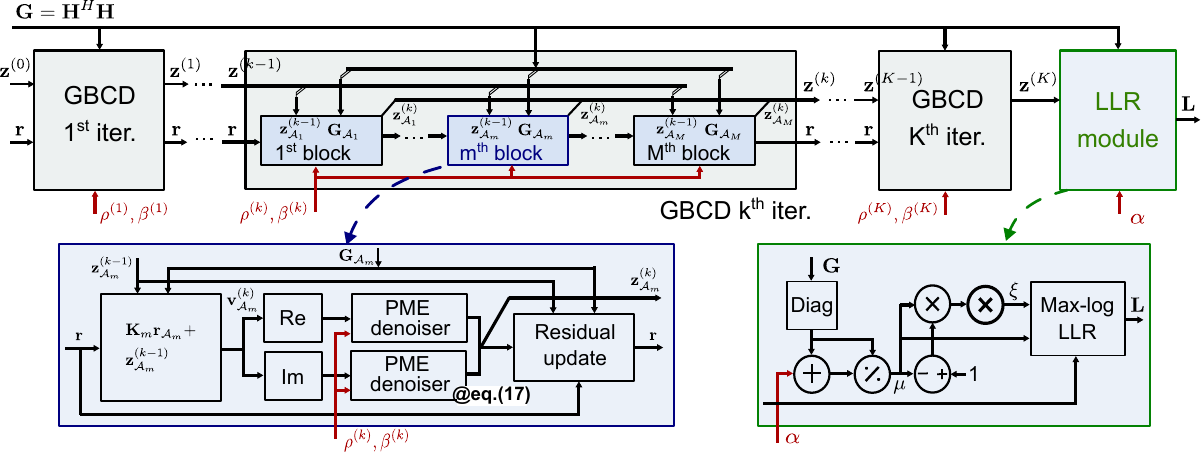}
    \caption{General deep unfolded architecture of the GBCD algorithm with the piecewise linear approximate PME denoiser.}
    \label{fig:deep_unfold_arch}
\end{figure*}

An exact analytical computation of the parameters $\rho$ and $\beta$ required in \fref{eq:PMEapprox} is challenging. To address this, the work in~\cite{Hu11} proposes to calculate the slope $\rho$ through differentiation of the original PME function for a similar piecewise linear approximation, in the context of estimate-and-forward relays. However, this approach still requires the exact value of $\omega$ in \fref{eq:explicitPME}, which makes this approach not suitable for hardware implementation. Therefore, we utilize deep unfolding~\cite{song21, sun24}, which leverages machine learning for automated and offline parameter tuning. The overall deep-unfolding architecture of GBCD with the piecewise linear approximate PME denoiser is shown in \fref{fig:deep_unfold_arch}, omitting Gram computation, matched filtering, and UE sorting, as these modules do not require any parameter tuning. To be specific, we unfold all inner and outer iterations of GBCD into a neural-network-like structure, and set $\rho^{(k)}$ and~$\beta^{(k)}$ as trainable parameters --- all UE blocks use the same set of parameters $\rho^{(k)}$ and $\beta^{(k)}$ in the $k$th iteration, as training different parameters for each UE block did not lead to any performance advantage. 
Additionally, we treat the normalization parameter $\alpha$ introduced in \fref{eq:LLRmu} for LLR computation as a trainable parameter. Therefore, for a $K$-iteration GBCD-PME algorithm, the model comprises $2K+1$ trainable parameters.

We next detail the employed training loss function. We first convert the bitwise LLR outputs in \fref{eq:LLR} into bit probabilities with the following well-known relation
\begin{equation} \label{eq:LLR2P}
    P_{u,b}= \textstyle \frac{1}{2}\!\left(1+\operatorname{tanh}\!\left(\frac{1}{2} \textit{LLR}_{u,b}\right)\right)\!,
\end{equation}
where $P_{u,b}$ represents the probability that the $b$th bit of the symbol transmitted by the $u$th UE is equal to 1, and $\textit{LLR}_{u,b}$ is the LLR value for the same bit. 
In what follows, $X_{u,b}$ denotes the binary value of the $b$th bit transmitted by the $u$th UE.
The training loss for a single transmission is given by the binary cross-entropy (BCE) 
\begin{equation}
    \textit{Loss} = \sum_{u=1}^{U} \! \sum_{b=1}^{\log_2(Q)} \!\!\! -(X_{u,b}\log(P_{u,b}) + (1-X_{u,b})\log(1-P_{u,b})).
    \label{eq:loss}
\end{equation}
This loss is used to train the model parameters offline using stochastic gradient descent, facilitated by the Adam optimizer~\cite{kingma2017adam}.
To prevent overfitting, we incorporate an early stopping mechanism with a window size of $10$ into the training system, i.e., training terminates if the loss on the validation set does not drop for $10$ consecutive epochs. 

To accommodate for different communication scenarios, we train distinct parameter sets for each relevant SNR, modulation, and channel propagation conditions (i.e., LoS or non-LoS). For each scenario, we generate 10\,000 training samples. Each sample consists of a unique channel matrix, a transmit vector, and the corresponding receive vector. The validation set is of the same size as the training set but generated using different channels. The batch size is set to be 100. The training SNR range is from $0$\,dB to $25$\,dB. For data detection with SNRs exceeding $25$\,dB, the same parameters trained at $25$\,dB are used. If the receive SNR is below $0$\,dB, then the BOX denoiser is used to attain robust performance.

\begin{algorithm}[t]
\caption{Gram Block Coordinate Descent (GBCD) Algorithm for Soft-Output Massive MU-MIMO Data Detection}\label{alg:GBCD}
\begin{algorithmic}[1]
\STATE \textbf{inputs:} $\bH$, $\bmy$, and \No\\
\STATE \textbf{parameters:} $\mathcal{O}$, $Q$, $L$, $K$, \texttt{mode}\\
\STATE \textbf{parameters trained offline via deep unfolding:}\\
\STATE $\bm{\omega}=[\omega^{(1)},\ldots,\omega^{(K)}]^{T}$, $\bm{\beta}=[\beta^{(1)},\ldots,\beta^{(K)}]^{T}$, $\alpha$ \\
\STATE \textbf{preprocessing: }\\
\STATE $\bG=\bH^\Herm\bH$, $\bmz^{(0)} = \bm{0}^{U\times 1}$\\
\FOR{$u=1$ to $U$} 
    \STATE $\lambda_u=\sum_{i=1, i\neq u}^{U}|G_{u,i}|^2$ \\
    {$\textit{SINR}_u^{-1} = \frac{\lambda_u}{|G_{u,u}|^2}+\frac{\No}{\Es |G_{u,u}|}$}\\
\ENDFOR
\STATE Sort elements of $\{\textit{SINR}^{-1}_u\}$ in ascending order and generate the corresponding index vector $\bm{\nu}=[\nu_1,\ldots,\nu_U]$ \\
\FOR{$m=1$ to $U/L$}
    \STATE $\mathcal{A}_m = \{\nu_{(m-1)L+1},\ldots,\nu_{mL}\}$\\
    \STATE $\bK_m = (\bG_{\mathcal{A}_m,\mathcal{A}_m})^{-1}$\\
\ENDFOR
\STATE \textbf{equalization: }\\
\STATE $\bmy^{\mathrm{MF}}=\bH^\Herm\bmy$, $\bmr=\bmy^{\mathrm{MF}}$\\
\FOR{$k=1$ to $K$}
    \FOR{$m=1$ to $U/L$}
        \STATE $\bmv_{\mathcal{A}_m}^{(k)} = \bK_m \bmr_{\mathcal{A}_m} + \bmz_{\mathcal{A}_m}^{(k-1)}$\\
        \IF{\texttt{mode} is BOX}
            \STATE $\bmz_{\mathcal{A}_{m}}^{(k)}=\mathcal{B}_{\mathcal{O}}(\bmv_{\mathcal{A}_{m}}^{(k)})$\\ 
        \ELSIF{\texttt{mode} is PME}
            \STATE $\bmz_{\mathcal{A}_{m}}^{(k)}=\tilde{\mathcal{P}}_{\mathcal{O}}(\bmv_{\mathcal{A}_{m}}^{(k)};\rho^{(k)},\beta^{(k)})$ \\ 
        \ENDIF
        \STATE $\Delta \bmz_{\mathcal{A}_{m}}^{(k)} = \bmz_{\mathcal{A}_{m}}^{(k)} - \bmz_{\mathcal{A}_{m}}^{(k-1)}$\\
        \STATE $\bmr\leftarrow \bmr - \bG_{\mathcal{A}_{m}}\Delta \bmz_{\mathcal{A}_{m}}^{(k)}$ \\
    \ENDFOR
\ENDFOR
\STATE $\hat{\bms} = \bmv^{(K)}$\\
\STATE \textbf{LLR outputs:}\\
\FOR{$u=1$ to $U$}
    \STATE $\mu_u=\frac{G_{u,u}}{G_{u,u}+\alpha}$, $\xi_u = E_s(1-\mu_u)\mu_u$\\
    \FOR{$b=1$ to $\log_2(Q)$}
       \STATE  $\textit{LLR}_{u,b}=$\\
        $\frac{1}{\xi_{u}}(\min_{a\in \mathcal{O}_{b}^0}|\hat{s}_{u}-\mu_u a|^2-\min_{a\in \mathcal{O}_{b}^1}|\hat{s}_{u}-\mu_u a|^2)$\\
    \ENDFOR
\ENDFOR
\end{algorithmic}
\end{algorithm}

We summarize the proposed GBCD-PME algorithm, along with the GBCD-BOX algorithm, in \fref{alg:GBCD}. As will be noted in \fref{sec:performance}, the GBCD-PME algorithm with trained parameters provides a notable SNR gain over the GBCD-BOX algorithm with only little hardware implementation overhead.

\subsection{Complexity Comparison}
\label{sec:complexity}

\begin{figure}[t]
	\centering
	\includegraphics[width=0.8\columnwidth]{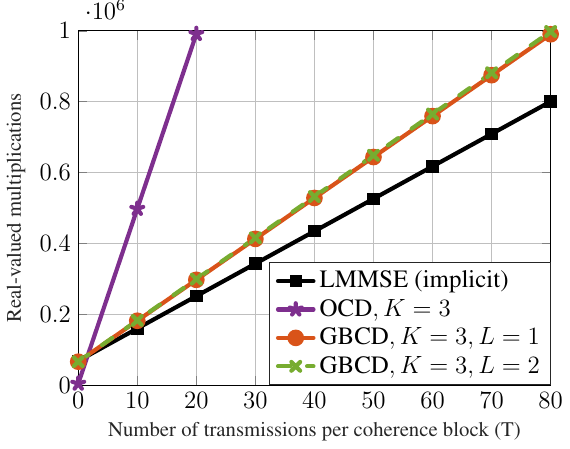}
	\caption{Complexity, measured in terms of the number of real-valued multiplications of the considered algorithms, versus the number of transmissions~$T$ per coherence block. }
	\label{fig:MulversusT}
\end{figure}

We now investigate the complexity of GBCD, OCD, and implicit LMMSE data detection~\cite{wu17a} by counting their required number of real-valued multiplications. {For simplicity, we assume that multiplication and division have the same complexity.}
Throughout the paper, we also assume that the preprocessing quantities are computed once per channel realization and reused over a block of $T$ transmissions, while equalization is performed for each receive vector~$\bmy$.
Here, we consider implicit LMMSE equalization as it has the lowest complexity among all LMMSE implementations~\cite{wu17a}, which is achieved by avoiding direct matrix inversion; instead, the solution vector is found via a Cholesky decomposition of the regularized Gram matrix followed by forward and backward substitution.
\fref{fig:MulversusT} shows an example comparison for a system with $B=128$ BS antennas and $U=16$ UEs. The y-intercept of each line in \fref{fig:MulversusT} corresponds to the preprocessing complexity.

For the proposed GBCD detector with block size $L=2$, the preprocessing phase requires $2BU^2+U(2U+2)+3U$ real-valued multiplications to compute the Gram matrix, SINR, and $2 \times 2$ matrix inversions. For each receive vector, equalization consists of matched filtering followed by $K$ GBCD iterations, which amounts to $4BU+8KU+4KU^2$ real-valued multiplications.
\fref{fig:MulversusT} shows that GBCD with block size $L=2$ does not incur any noticeable complexity increase compared to GBCD with block size $L=1$, while as shown in \fref{sec:individual_effects}, when combined with UE sorting, setting the block size to $L=2$ significantly outperforms  $L=1$. Furthermore, GBCD with $L=2$ has almost half the latency of GBCD with $L=1$.

Compared to OCD, which requires low preprocessing complexity but higher equalization complexity, incorporating the Gram-domain transformation in the preprocessing stage of GBCD enables more than $3\times$ reduction in complexity for \mbox{$T>10$}. 
Preprocessing of implicit LMMSE equalization requires $2BU^2+2/3U^3 - 2/3U$ real-valued multiplications, where $2/3U^3 - 2/3U$ is for the Cholesky decomposition. Despite the slightly higher preprocessing complexity compared to GBCD, implicit LMMSE equalization achieves lower complexity as $T$ increases.
However, the highly sequential nature of the implicit LMMSE equalizer not only incurs high latency, but also prevents the efficient reuse of hardware resources. Additionally, as shown in \fref{sec:performance}, GBCD with the PME denoiser can outperform LMMSE equalization in terms of error~rate for channels exhibiting strong correlation. 

\subsection{Coded Block Error-Rate (BLER) Performance}
\label{sec:performance}

\begin{figure*}[t]
	\centering
	\subfigure[$128\times 16$, $R=5/6$, non-LoS, $256$-QAM]
	{\includegraphics[width=0.45\textwidth]{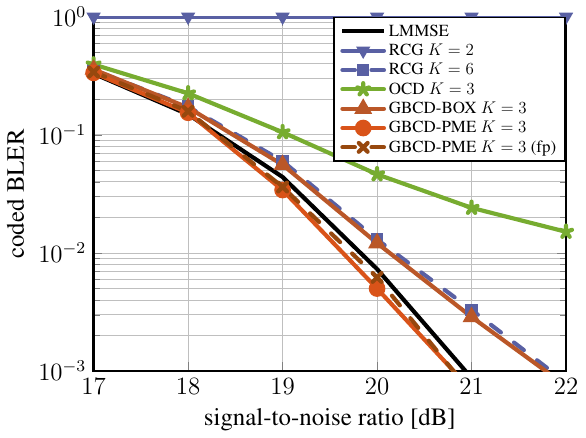}\label{fig:PERNLOS256QAM}}
	\hfill
	\subfigure[$128\times 16$, $R=3/4$, LoS, $256$-QAM]
	{\includegraphics[width=0.45\textwidth]{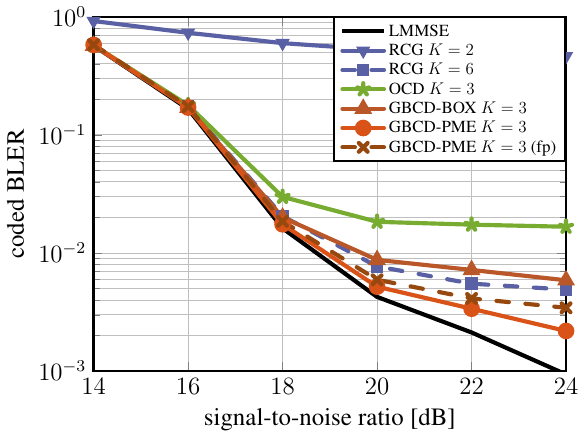}\label{fig:PERLOS256QAM}}
	\subfigure[$16\times 16$, $R=1/2$, non-LoS, QPSK]
	{\includegraphics[width=0.45\textwidth]{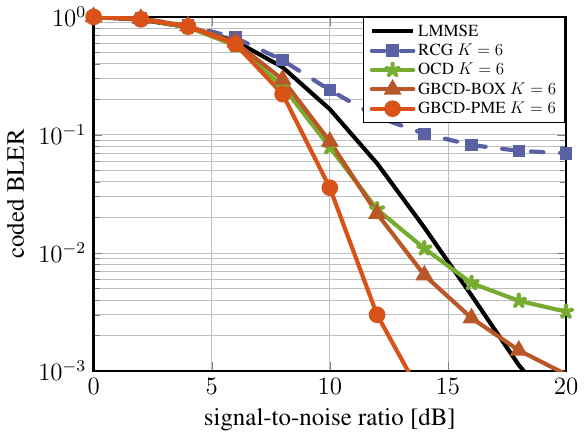}\label{fig:PERNLOSQAM1616}}
	\hfill
	\subfigure[$16\times 16$, $R=1/2$, LoS, QPSK]
	{\includegraphics[width=0.45\textwidth]{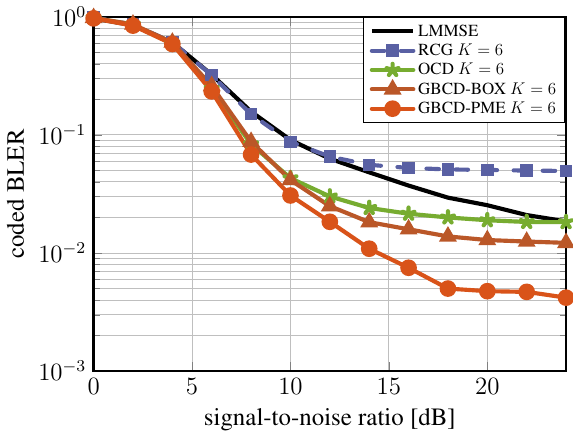}\label{fig:PERLOSQAM1616}}
	
	\caption{Coded BLER simulation results for a $128 \times 16$ system with 256-QAM data in (a) non-LoS and (b) LoS channels, demonstrates that the proposed GBCD-PME with three iterations achieves better than or near LMMSE equalization performance in the practical BLER range, while RCG suffers a notable performance loss compared to LMMSE equalization even with six iterations. The coded BLER of a $16 \times 16$ system with QPSK data in (c) non-LoS and (d)~LoS channels, indicates that in high load scenarios the proposed GBCD-PME can significantly outperform LMMSE equalization. The curves with ``(fp)'' in their labels correspond to the fixed-point performance of the implemented ASIC.}
	\label{fig:BLER}
\end{figure*}

To compare the performance of GBCD with both BOX and approximate PME denoisers against reference data detectors that are suitable for hardware implementation, we simulate a massive MU-MIMO-OFDM system with $N=2048$ subcarriers, out of which $N_d=1200$ carry data. Each UE employs a punctured rate-$R$ convolutional code with constraint length~$7$, generator polynomials $[133_o, 171_o]$, and random interleaving. Each UE encodes a block of $N_d \log_2(Q) R$~bits into $N_d \log_2(Q)$~coded bits that are then mapped into one OFDM symbol with $N_d$ data-carrying subcarriers.

We use QuaDRiGa mmMAGIC~\cite{QuaDRiGa_tech_rpt} to generate urban microcell (UMi) LoS and non-LoS mmWave channel vectors at a carrier frequency of $60$\,GHz. This simulator generates realistic multipath fading channel realizations for the specified settings.
We also emulate power control to ensure that the variation in the receive power of the UEs is limited to $\pm 3$\,dB. We assume that the BS has access to perfect channel state information (CSI). However, as demonstrated in \fref{sec:chest}, the same error-rate performance trends are observed with estimated CSI.

In \fref{fig:BLER}, we show the coded BLER for both $128\times 16$ (which stands for $B=128$ and $U=16$) and $16\times 16$ MU-MIMO systems. For each system setting, we set the code rate and modulation order to make it possible to achieve a coded BLER of at least $1\%$ in an SNR less than $20$\,dB, as higher SNRs are not practical.
Adapting the code rate and modulation scheme based on channel conditions, known as modulation coding scheme (MCS) adaptation, is done in virtually all practical wireless systems, including 5G new radio \cite{TS38_214_v18}.

For the $128\times 16$ system, we employ 256-QAM data with a code rate of $R=3/4$ for LoS channels and $R=5/6$ for non-LoS channels. 
For the non-LoS channels in \fref{fig:PERNLOS256QAM}, GBCD-PME outperforms LMMSE equalization by $0.2$\,dB SNR at $1$\% BLER. Furthermore, both GBCD-PME and GBCD-BOX significantly outperform OCD, which suffers from a high error floor. 
Grouping UEs into blocks, UE sorting, and deep-unfolding-assisted PME denoising, all contribute to successful data detection.
The individual impact of each of these techniques is explored in \fref{sec:individual_effects}.

For the highly correlated mmWave LoS channels in \fref{fig:PERLOS256QAM}, LMMSE equalization outperforms GBCD for target BLERs lower than $1$\%. 
However, in most modern wireless communication applications, such as LTE and 5G NR, a target coded BLER of $10$\% to $1$\% is practically relevant~\cite{3gpp22, Chu19}, where GBCD-PME performs virtually on par with LMMSE equalization. 
Finally, we observe that the recursive conjugate gradient (RCG) detector proposed in \cite{liu20} results in a high error floor in correlated channels with imperfect power control. This is particularly evident for the implemented RCG ASIC, which only performs $K=2$ algorithm iterations. We also include the performance of RGC with $K=6$ iterations to show that for such channels, RCG needs more iterations to achieve an acceptable performance. 

In Figures~\ref{fig:PERNLOSQAM1616} and~\ref{fig:PERLOSQAM1616}, we show simulation results for a
more challenging $16\times 16$ system, where we employ a code rate of $R=1/2$ and $K=6$ iterations with QPSK modulation. 
These results show that GBCD-PME achieves a significant SNR gain compared to LMMSE equalization of $4$\,dB for non-LoS channels and more than $10$\,dB for LoS channels at $1$\%~BLER. While the VLSI design discussed in \fref{sec:vlsi} is not tailored to this setup with $16$ BS antennas, our design can easily be adapted to such systems.

\subsection{Individual Impact of Each Technique}
\label{sec:individual_effects}

\begin{figure}[t]
	\centering
	\subfigure[$128\times 16$, $R=5/6$, non-LoS, $256$-QAM]
	{\includegraphics[width=0.45\textwidth]{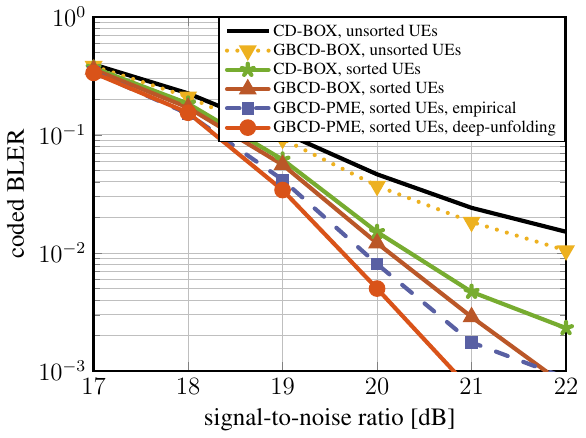}\label{fig:module_nonlos}}
	\hfill
	\subfigure[$128\times 16$, $R=3/4$, LoS, $256$-QAM] 
	{\includegraphics[width=0.45\textwidth]{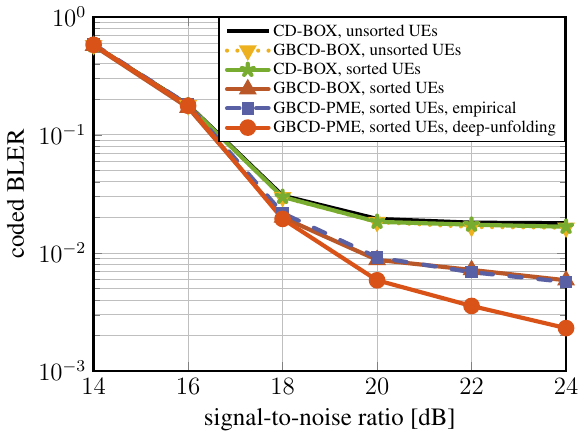}\label{fig:module_los}}
	\caption{Coded BLER simulation results with each proposed technique applied incrementally to the original BOX-based coordinate descent detector, OCD.}\label{fig:modules}
\end{figure}

Although \fref{fig:BLER} shows the performance improvement achieved by the proposed GBCD-PME algorithm compared to the baseline methods, the individual impact of each of the proposed techniques, i.e., UE sorting, grouping the UEs into blocks via BCD, PME denoiser, and deep-unfolding-assisted parameter tuning for PME, is not shown. 
To investigate the effect of each one of these techniques, we start from the original OCD, and incrementally apply each technique to finally arrive at the proposed deep-unfolding-assisted GBCD-PME.

The results of this experiment for the $128 \times 16$ system are shown in \fref{fig:modules}, where the legends indicate the techniques. The starting point is the original OCD, which corresponds to CD with the BOX denoiser (thus, the curves for ``OCD'' in \fref{fig:BLER} have been renamed to ``CD-BOX'' in \fref{fig:modules}). Since the Gram-domain transformation does not change the system model and has no effect on the performance, BCD and the Gram-domain transformation are applied at the same time (indicated by ``GBCD'' in \fref{fig:modules}). In order to analyze the impact of deep-unfolding-assisted parameter tuning for PME, as an alternative, we performed empirical parameter tuning by searching for the optimal PME parameters over a grid. Note that since there are $2K+1$ parameters to be optimized (including one for the LLR unit), an extensive grid search over all of them is infeasible. Hence, we use a single pair of PME parameters for all GBCD iterations and find the empirical optimal value with this simplified grid search. The result is labeled as ``empirical'' in \fref{fig:modules}.

 \fref{fig:modules} shows that BCD alone does not improve the error-rate performance notably.
Likewise, the proposed SINR-based UE sorting alone does not improve the error error-rate performance for LoS channels (\fref{fig:module_los}), although it has a significant impact for non-LoS channels (\fref{fig:module_nonlos}).
However, combining BCD with UE sorting results in a significant performance improvement for both non-LoS and LoS conditions: In non-LoS channels, for example, we observe a reduction of more than $1.7$\,dB in the SNR required to achieve $1\%$ BLER compared to the original OCD (i.e., CD-BOX with unsorted UEs).
The PME with empirically tuned parameters improves over the BOX denoiser for non-LoS channels, but not for LoS channels.
Nevertheless, using the deep-unfolding-assisted PME denoiser outperforms the BOX denoiser by $0.5$\,dB under both propagation conditions.
These results demonstrate that all of the proposed techniques contribute to the excellent performance of the proposed GBCD-PME algorithm.

\subsection{Impact of Channel Estimation Errors}
\label{sec:chest}

\begin{figure}[t]
	\centering
	\subfigure[$128\times 16$, $R=5/6$, non-LoS, $256$-QAM]
	{\includegraphics[width=0.45\textwidth]{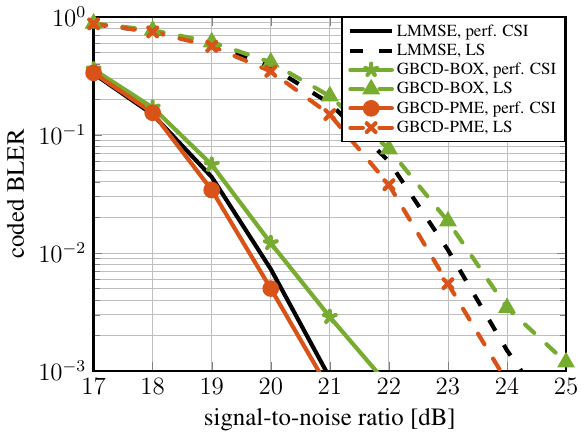}}
	\hfill
	\subfigure[$128\times 16$, $R=3/4$, LoS, $256$-QAM]
	{\includegraphics[width=0.45\textwidth]{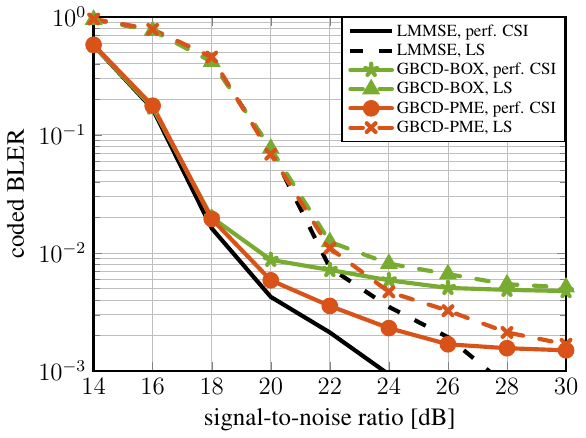}}
	\caption{Coded BLER performance under imperfect channel state information.}
	\label{fig:imperfcsi}
\end{figure}

So far we have assumed that the BS has access to perfect CSI. In order to investigate the impact of channel-estimation errors on the proposed algorithm, we now consider least squares (LS) channel estimation. LS channel estimates using orthogonal pilot sequences can be modeled as
$\hat{\bH} = \bH + \bE$,
where $\bE$ is an estimation error matrix with i.i.d.\ circularly symmetric complex Gaussian entries with variance~$\frac{N_0}{E_sU}$ per entry.

\fref{fig:imperfcsi} shows the coded BLER simulation results with perfect CSI and LS channel estimates as modeled above for the LMMSE, GBCD-BOX, and GBCD-PME algorithms for a $128 \times 16$ system with $256$-QAM signaling. All other simulation settings are the same as in \fref{sec:performance}. We observe that the proposed GBCD-PME algorithm exhibits the same relative performance compared to the LMMSE and GBCD-BOX data detectors, in both LoS and non-LoS channels. Therefore, we conclude that the proposed GBCD-PME algorithm maintains its relative performance advantages to other methods in the case of estimated channel matrices.\\

\section{\textsc{VLSI Architecture}}
\label{sec:vlsi}
We now present a VLSI architecture for the proposed GBCD detector with $K=3$ iterations for a $128\times 16$ massive MU-MIMO system which supports QPSK to 256-QAM. 
The VLSI architecture features a reconfigurable array of processing elements (PEs) and pipeline-interleaving, which maximizes utilization of the PEs and achieves high throughput at low area. 
In what follows, ``GBCD'' refers to the entire data detector, while ``BCD equalizer'' specifically refers to the module implementing the equalization parts from lines~17 to~36 in \fref{alg:GBCD}.

\subsection{Architecture Overview}
\label{sec:Archoverview}

\begin{figure}[t]
	\centering
	\includegraphics[width=1.0\columnwidth]{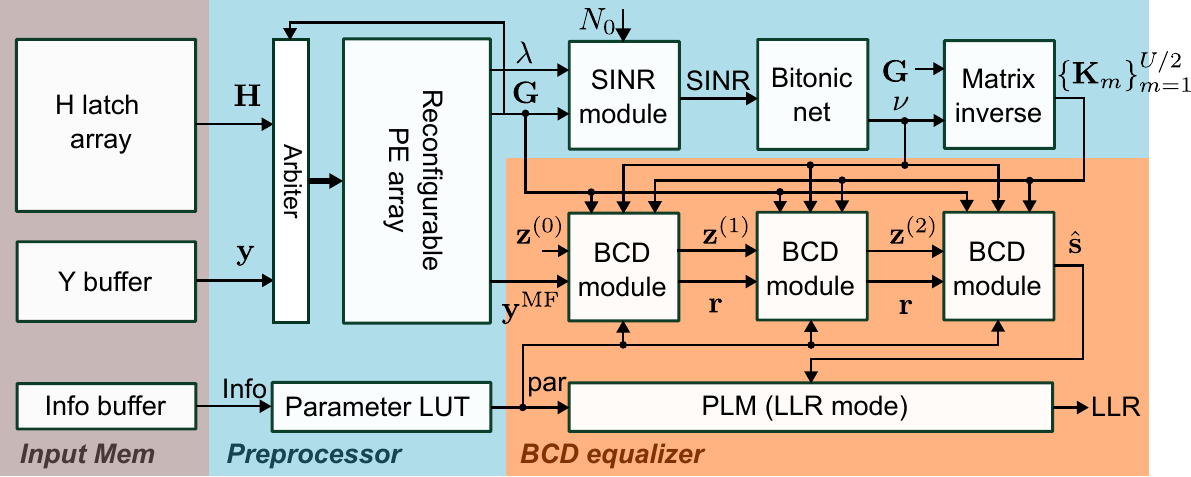}
	\caption{Top-level architecture of the proposed GBCD detector, consisting of  input memories,  a preprocessor, which performs preprocessing tasks as well as MF, and the BCD equalizer, which performs three iterations of the GBCD algorithm, each within a BCD module, and LLR computation.}
	\label{fig:top_level}
\end{figure}

\fref{fig:top_level} depicts the top-level architecture of GBCD, comprising three key components: input memory, preprocessor, and BCD equalizer.

The input memory stores the channel matrix $\bH$ in a latch array and buffers the incoming receive vectors~$\mathbf{y}$ in a register array. As detailed in the following subsection, different modes of operation in the array of PEs require different access patterns into $\bH$, with the most data-intensive mode reading out eight rows of $\bH$ per clock cycle. Compared to on-chip SRAM macrocells, the use of a latch array offers more flexible data access during the preprocessing phase. 

In the preprocessor, the reconfigurable PE array receives~$\bH$ and $\bmy$ to compute the Gram matrix $\bG = \bH^\Herm \bH$ and the MF vector $\bmy^{\mathrm{MF}} = \bH^\Herm \bmy$, as well as the interference terms $\lambda_{u}=\sum_{i=1,i\neq u}^{U}|G_{u,i}|^2$ in a time-multiplexed manner. With the derived interference term, the SINR is then computed by the SINR module and sent into a bitonic sorting network~\cite{Liang2021} for UE sorting. Finally, the inverses of $\frac{U}{2}$ $2 \times 2$ submatrices of the Gram matrix are computed in sequence within the same $2 \times 2$ matrix inverse module.

The BCD equalizer comprises three identical BCD modules and one LLR module to compute soft outputs. Each BCD module takes charge of one outer iteration of \fref{alg:GBCD}. The details of the reconfigurable PE array and the BCD modules are discussed in the following subsections.

\subsection{Reconfigurable PE Array}
\label{sec:reconfigure}

Most existing MIMO data-detector implementations that include preprocessing circuitry, such as those in \cite{peng18b,liu20}, separate the PEs for Gram and MF computation in order to optimize throughput. However, this strategy often leads to a long idle phase for the Gram module and low PE utilization, especially when the number of transmissions per coherence block $T$ is large. 
Furthermore, a dedicated preprocessor module incurs additional area overhead, which is impractical for OFDM-based multi-antenna receivers, as deploying hundreds of such parallel engines would result in prohibitively high silicon area. 
To address these issues, we introduce a reconfigurable PE array. This design optimizes PE utilization by employing the same PEs for Gram, MF, and interference term ($\lambda_u$ in line 8 of \fref{alg:GBCD}) computation during different time slots. This strategy increases utilization of the PE array to more than 90\%, and decreases the area for Gram and MF calculations by 12.5\% compared to the designs described in \cite{peng18b, liu20} by eliminating the need for two separate PE arrays.

\begin{figure}[t]
	\centering
	\includegraphics[width=0.9\columnwidth]{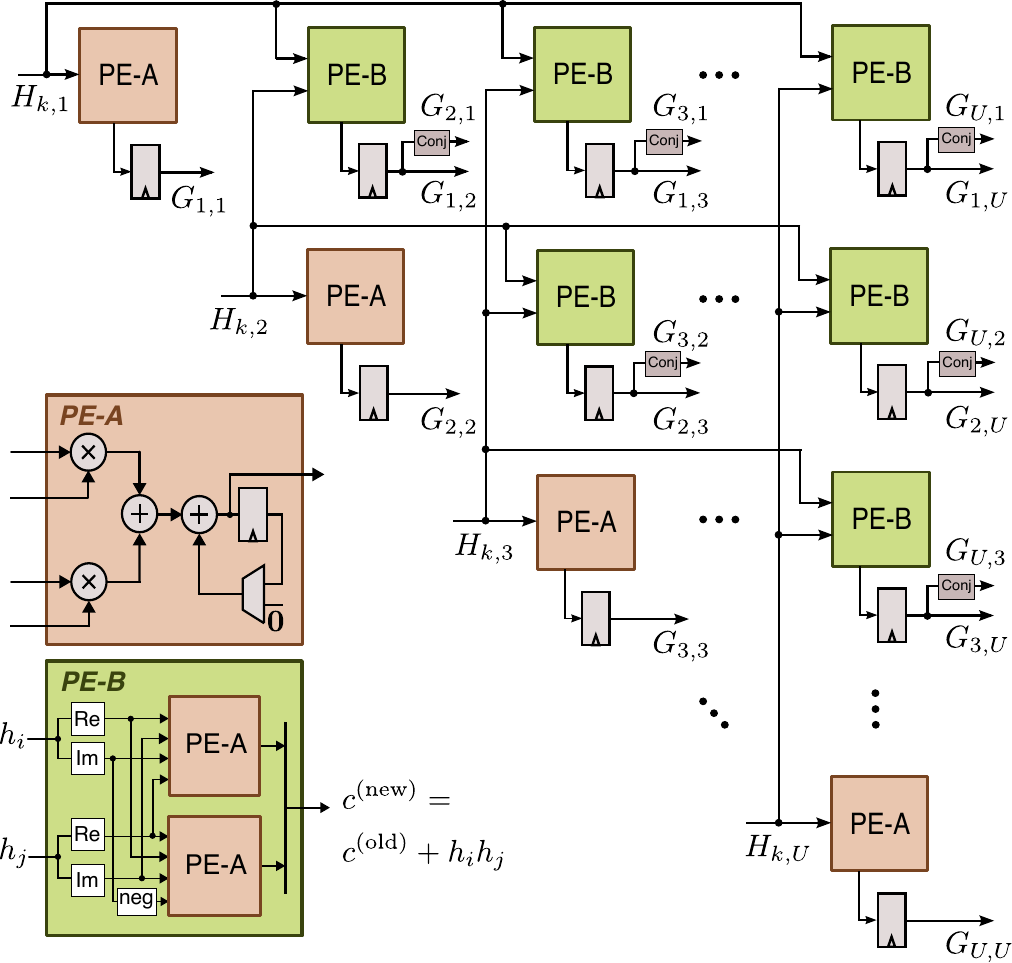}
	\caption{Architecture of the PE array in Gram mode (right) and internal architectures of PE-A and PE-B (left). The PE array computes the upper triangular part of $\bG$ in $B$ clock cycles and then derives the lower triangular part by conjugation. PE-A and PE-B are MAC units to compute diagonal and off-diagonal entries of $\bG$.}
	\label{fig:Gram_mode}
\end{figure}

The PE array is depicted in \fref{fig:Gram_mode} and comprises two types of units: PE-A and PE-B. PE-A performs a two-dimensional dot-product and accumulates the result.
When two PE-A units operate together, they form a complex-valued MAC unit, namely PE-B, as shown in the lower left of \fref{fig:Gram_mode}. 
The PE array includes $256$ PE-As, which can be reconfigured into at most $128$ PE-Bs depending on the operation mode. During the Gram mode, the PE array operates with $16$ PE-As and $120$ PE-Bs, while during the matched filtering mode, it is reconfigured as $128$ PE-B units.

The reconfigurability of the PE array is due to a dedicated arbiter that routes the data items to the PEs depending on the operation mode.

\subsubsection{Gram Computation}
During Gram computation, i.e., $\bG=\bH^\Herm\bH$ in line 6 of \fref{alg:GBCD}, the arbiter distributes the entries of $\bH$ to the PEs such that the PE array is organized as the upper triangular array shown in \fref{fig:Gram_mode}, with $U$ PE-As on the diagonal entries and $\frac{U(U-1)}{2}$ PE-Bs on the off-diagonal entries. The PE located on the $i$th row and $j$th column computes $G_{i,j}$ during $B=128$ clock cycles and the corresponding lower triangular entries are derived by conjugation.

\subsubsection{Matched Filtering}
\begin{figure}[tp]
	\centering
	\subfigure[MF mode]{\includegraphics[width=.65\columnwidth]{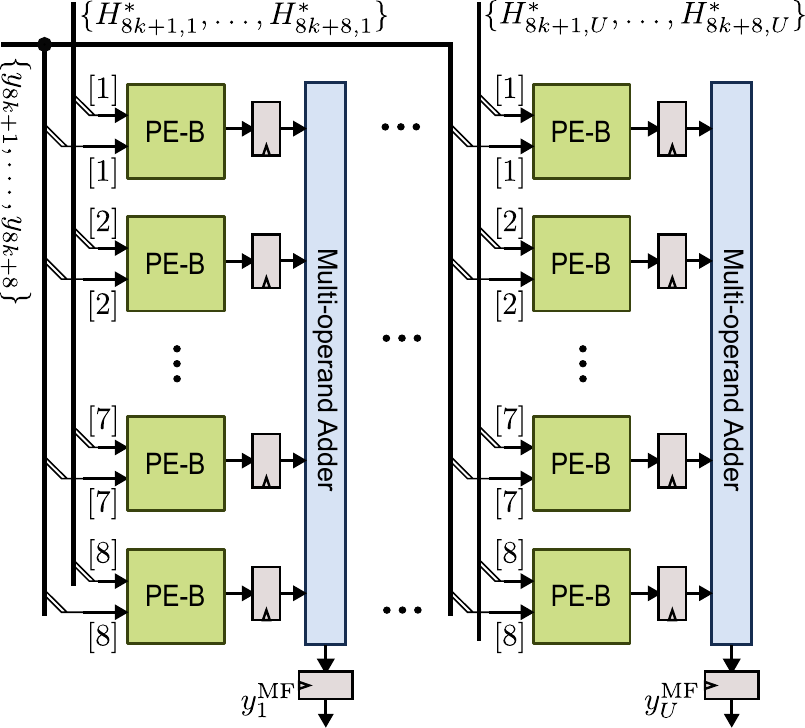}\label{fig:MF_mode}}
	\hfill
	\subfigure[Interference mode]{\includegraphics[width=.3\columnwidth]{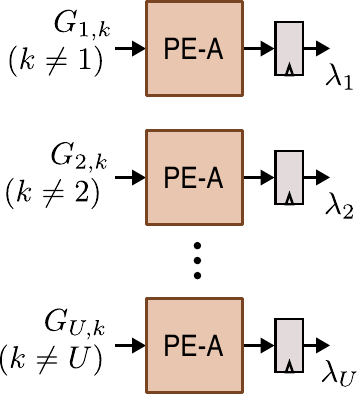}\label{fig:INT_mode}}
	\caption{Configuration of the PE array in MF and interference computation modes. In MF mode (a), the PE array is reconfigured as a $U 
	\times U/2$ array of PE-Bs which computes $\bmr = \bH^\Herm \bmy$ in $2B/U=16$ clock cycles. In SINR mode (b), $U=16$ PE-As compute the interference term of SINR in $15$ clock cycles.}
	\label{fig:PE_array}
\end{figure}

During the matched filtering operation shown in line 16, the PE array is reconfigured as a $U$-by-$\frac{U}{2}$ rectangular array of PE-Bs, as depicted in \fref{fig:MF_mode}.
The $i$th column of the array computes the $i$th entry of $\bmy^{\mathrm{MF}}$ with $U/2=8$ parallel PE-Bs:
\begin{align}
    y^{\mathrm{MF}}_{i}=\sum_{j=1}^{8}\sum_{k=1}^{16}H_{i,8(k-1)+j}^* \> y_{8(k-1)+j},
\end{align}
where the external summation is carried out by a multi-operand adder, and the $j$th internal summation is calculated by the PE-B located at $i$th column and $j$th row in $\frac{2B}{U}=16$ clock cycles.

\subsubsection{Interference Computation}
As illustrated in \fref{fig:INT_mode}, 16 PE-As inside the reconfigurable PE array operate in parallel to calculate the interference terms ($\lambda_u$'s in line 8) in the SINR, leaving the other PEs idle in this mode. The $u$th PE-A takes charge of computing $\lambda_{u}=\sum_{i=1,i\neq u}^{U}|G_{u,i}|^2$ in $U-1=15$ clock cycles. We emphasize that the interference computation is the only mode that leaves some PEs idle.

\subsection{SINR Computation, Sorting, and $2 \times 2$ Submatrix Inverse}
Concurrent to the computation of $\lambda_u$ values in the PE array, the scalars $a_u=1/|G_{u,u}|^2$ and $b_u=N_0/(E_s |G_{u,u}|)$ for all UEs are computed in $U$ clock cycles by the SINR module shown in \fref{fig:top_level}. Subsequently, this module computes $\textit{SINR}_u^{-1}=a_u\lambda_u+b_u$ for all UEs simultaneously in one additional clock cycle. The SINR computation, which corresponds to lines 7 to 9 of \fref{alg:GBCD}, is therefore completed in $U+1=17$ clock cycles. Then, the UE indices are sorted based on UEs' reciprocal SINR values in ascending order using a bitonic sorting network~\cite{Liang2021}, according to line 10 of \fref{alg:GBCD}. This sorting network is suitable for parallel processing and leads to a short delay of $\frac{1}{2}(\log\sb{2}U)(\log\sb{2}U+1)=10$ clock cycles.

After the sorting operation, the inverse of the $2 \times 2$ submatrices $\bG_{\setA_m,\setA_m}$, $m=1,2,\ldots,U/2$, are computed sequentially using a single $2 \times 2$ matrix inverse unit (lines~11 to~14 of \fref{alg:GBCD}).
Each submatrix inversion requires 2 clock cycles, leading to a total delay of $U=16$ clock cycles.

\subsection{BCD Module}
\label{sec:BCD_eq}

\begin{figure}[t]
\centering
\includegraphics[width=1.0\columnwidth]{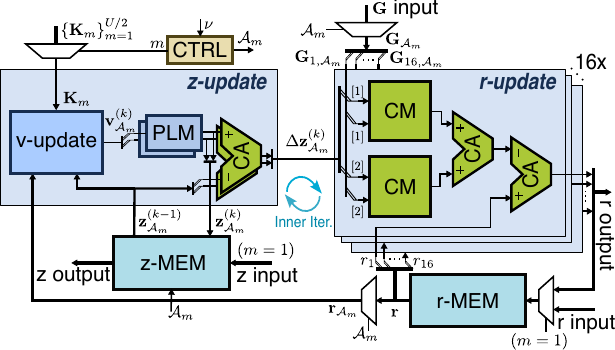}
\caption{Architecture of the BCD module, consisting of (i) z-update module to update the estimates $\bmz$, (ii) r-update module to update the residual term $\bmr$, and (iii) z-MEM and r-MEM to store $\bmz$ and $\bmr$.}
\label{fig:BCD_unit}
\end{figure}

The BCD module, depicted in \fref{fig:BCD_unit}, implements one inner iteration of \fref{alg:GBCD} from line 19 to 26. This module includes two sequential units: (i) the \emph{z-update} unit, which performs the operations corresponding to lines 19 to 25, and (ii) the \emph{r-update} unit that updates the residual vector according to line~26. This BCD module repeats these steps for $m=1,\ldots,U/2$ to complete one outer iteration of \fref{alg:GBCD} and then passes the results to the next BCD module.

A controller unit (CTRL) counts the inner iteration index~$m$, takes the sorted indices $\bm{\nu}$ from the preprocessor, and constructs the $m$th block index $\mathcal{A}_m$ accordingly. $\mathcal{A}_m$ and $m$ are then used in the BCD module as selectors of multiplexers to select the part of the matrices and vectors corresponding to the current inner iteration $m$.
Within the z-update module, the \emph{v-update} submodule contains multipliers and adders to compute the noisy estimates $\bmv_{\mathcal{A}_m}^{(k)}$ (line 19 of \fref{alg:GBCD}).
The symbol denoising step (line 20 to 24 of \fref{alg:GBCD}) is then performed by a piecewise linear mapping (PLM) module, which is detailed in the next subsection. 
Finally, the z-update module calculates $\Delta \bmz^{(k)}_{\mathcal{A}_{m}}$ with a complex-valued adder (CA) and overwrites $\bmz^{(k-1)}_{\mathcal{A}_{m}}$ with the new value $\bmz^{(k)}_{\mathcal{A}_{m}}$ in the memory module storing the $\bmz$ vector (z-MEM).

Next, two columns of $\bG$ indexed by $\mathcal{A}_m$ are fed into the r-update unit, which implements line 26 of \fref{alg:GBCD} in 16 parallel layers, each responsible for updating one entry of $\bmr$, using complex-valued multipliers (CMs) and CAs. Both the z-update and r-update units have a delay of one clock cycle. 
Therefore, the BCD module needs two clock cycles to perform one inner iteration of \fref{alg:GBCD} and $2 \times U/2 = 16$ clock cycles to complete one outer iteration. 
\subsection{Piecewise Linear Mapping (PLM) Module}
\label{sec:PLM}
\fref{fig:PLM module} shows the PLM module which implements the BOX and PME denoisers, as well as LLR computation, all with the same architecture. This module consists of LUTs that store the parameters needed to implement the function based on the selected mode.
The range identifier compares the input with pre-computed bin boundaries (stored in an LUT inside the range identifier) to derive the bin index. The bin boundaries are illustrated in the bottom of \fref{fig:PLM module} by vertical dashed lines.  
The index produced by the range identifier is then provided as address to the slope LUT and bias LUT to get the slope and bias of the corresponding bin.
Then, the affine mapping is achieved by multiplying the input with the slope and adding the bias.
We emphasize that the PLM parameters for different communication scenarios are precomputed offline 
based on trained values $\rho$ and $\beta$, and are stored in the parameter LUT depicted in \fref{fig:top_level}. During preprocessing, the system parameters (i.e., modulation order, SNR and LoS or non-LoS channel condition) are encoded as an address for the parameter LUT. The output values, denoted as ``par'' in \fref{fig:top_level}, are distributed to the PLM modules and are used to initialize the LUTs inside PLMs and the LLR module.

By implementing the approximate PME discussed in \fref{sec:PMEapprox} using the PLM module, our design achieves the following advantages over the PME implementation in \cite{jeon19b}: (i)~a unified datapath applicable to all modulation orders, and (ii) a significantly lower latency in each inner iteration of \fref{alg:GBCD}, which directly translates to higher throughput. 
Moreover, the PLM module can be configured to implement the BOX denoiser, which provides a reliable alternative in extreme conditions lacking ML-trained parameters for PME. 

\subsection{LLR Module}
\begin{figure}[t]
\centering
\includegraphics[width=1.0\columnwidth]{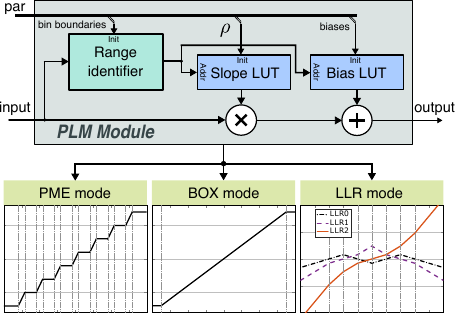}
\caption{Architecture of the piecewise linear mapping (PLM) module.}
\label{fig:PLM module}
\end{figure}

The LLR module in \fref{fig:top_level} is implemented by the PLM module in LLR mode as outlined in \fref{fig:PLM module}. Not shown in \fref{fig:PLM module} is the additional circuitry of the LLR module required to compute channel gain and NPI variance (cf. line~32 of \fref{alg:GBCD}). This additional circuitry comprises adders, multipliers, and a reciprocal LUT for calculating $\frac{1}{G_{u,u}+\alpha}$.

\subsection{Timing Schedule}
\label{sec:timing}

\begin{figure}[tp]
\centering
\includegraphics[width=1.0\columnwidth]{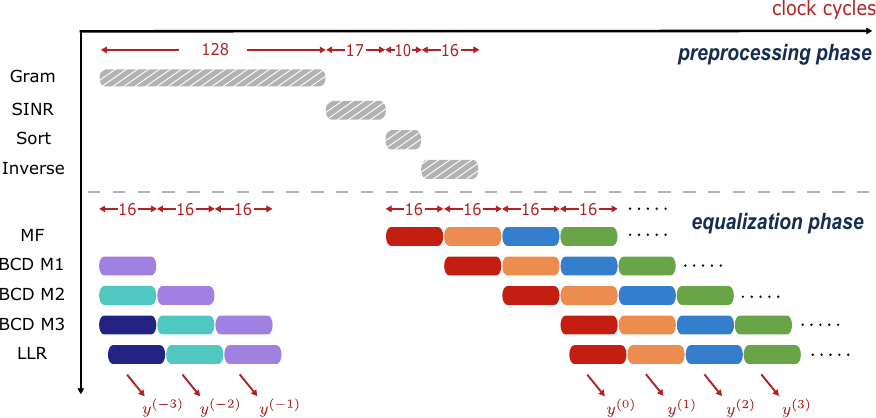}
\caption{Time schedule of the proposed GBCD detector. The gray bars with diagonal lines represent the preprocessing tasks, while the coloured bars represent the equalization tasks. Varied colours denote the processing flows for different receive vectors.}
\label{fig:time_schedule}
\end{figure}

Given the latency of each module, the timing schedule of the proposed architecture is shown in \fref{fig:time_schedule}. The gray hatched bars represent the preprocessing phase, while the colored bars illustrate the equalization phase, with different colors representing different receive vectors. Note that the receive vectors with negative superscripts are the transmissions from the previous coherence block. 

To achieve maximum throughput in the equalization phase, we process four successive receive vectors, e.g., $\bmy^{(t)}$ to $\bmy^{(t+3)}$, simultaneously in a pipeline-interleaved manner. After $\bmy^{(t)}$ undergoes MF, we move it into the first BCD module and start MF for $\bmy^{(t+1)}$. We repeat this process to ensure every module is fully occupied. Since the delay of MF and each BCD module is equal to 16 cycles for the target $128\times 16$ system, this approach leads to full hardware utilization when four data vectors have consecutively flushed in. 
Due to this processing scheme, a new equalized vector is produced every 16~clock cycles.

We emphasize that the PE array needs to pause the MF when it is computing the Gram matrix and interference term of SINR for a new channel due to the shared reconfigurable PE array. This resource conflict results in an idle phase of $B+U=144$ clock cycles, during which no new data detection task can start. 
Thus, the total number of clock cycles for accomplishing $T$ detection tasks including the preprocessing phase is $16T+144$ and the throughput of the GBCD detector is therefore given by
\begin{align} \label{eq:throughput}
    \Theta(T) = \frac{T}{16T+144} \log_2(Q)U  f_{\mathrm{clk}} ~~~[\mathrm{bps}].
\end{align}

The BCD equalizer is active for $16T$ cycles. We define the equalizer utilization as:
\begin{align} \label{eq:BCDutilization}
	\eta_\text{equ} = \frac{16T}{16T+144}=\frac{T}{T+9},
\end{align}
which will serve as an important parameter in \fref{sec:ASICMeasure}.

After careful pipelining, the critical path of the design goes through the z-update module and contains two real-valued multipliers and four real-valued adders. 

\subsection{Fixed-Point Design Parameters}
To maximize area and energy efficiency, we exclusively use fixed-point arithmetic.
For all complex-valued signals, we use the same number of bits for each real and imaginary part, so we only report the number of bits for one of these parts in what follows.
The inputs $\bH$ and~$\bmy$ are both quantized to 12 bits,  
while the preprocessor outputs $\bG$ and $\bmy^{\text{MF}}$ are represented with 15 and 18 bits, respectively.
The symbol estimate~$\bmz$ and the real-valued LLR outputs are quantized to 11 and 18 bits, respectively. The increased precision for some key quantities (e.g., $\bG$ or $\bmy^{\text{MF}}$) is necessary to support higher-order constellations. The results labeled with ``(fp)'' in \fref{fig:BLER} show that, at a coded BLER of $10^{-2}$ with 256-QAM, the fixed-point SNR loss is less than $0.1$\,dB for both mmWave non-LoS and LoS channels. 
For lower-order constellations, the SNR loss due to fixed-point arithmetic is even smaller.

\section{\textsc{ASIC Implementation Results}}
\label{sec:asic}

\begin{figure}[t]
\centering
\includegraphics[width=0.9\columnwidth]{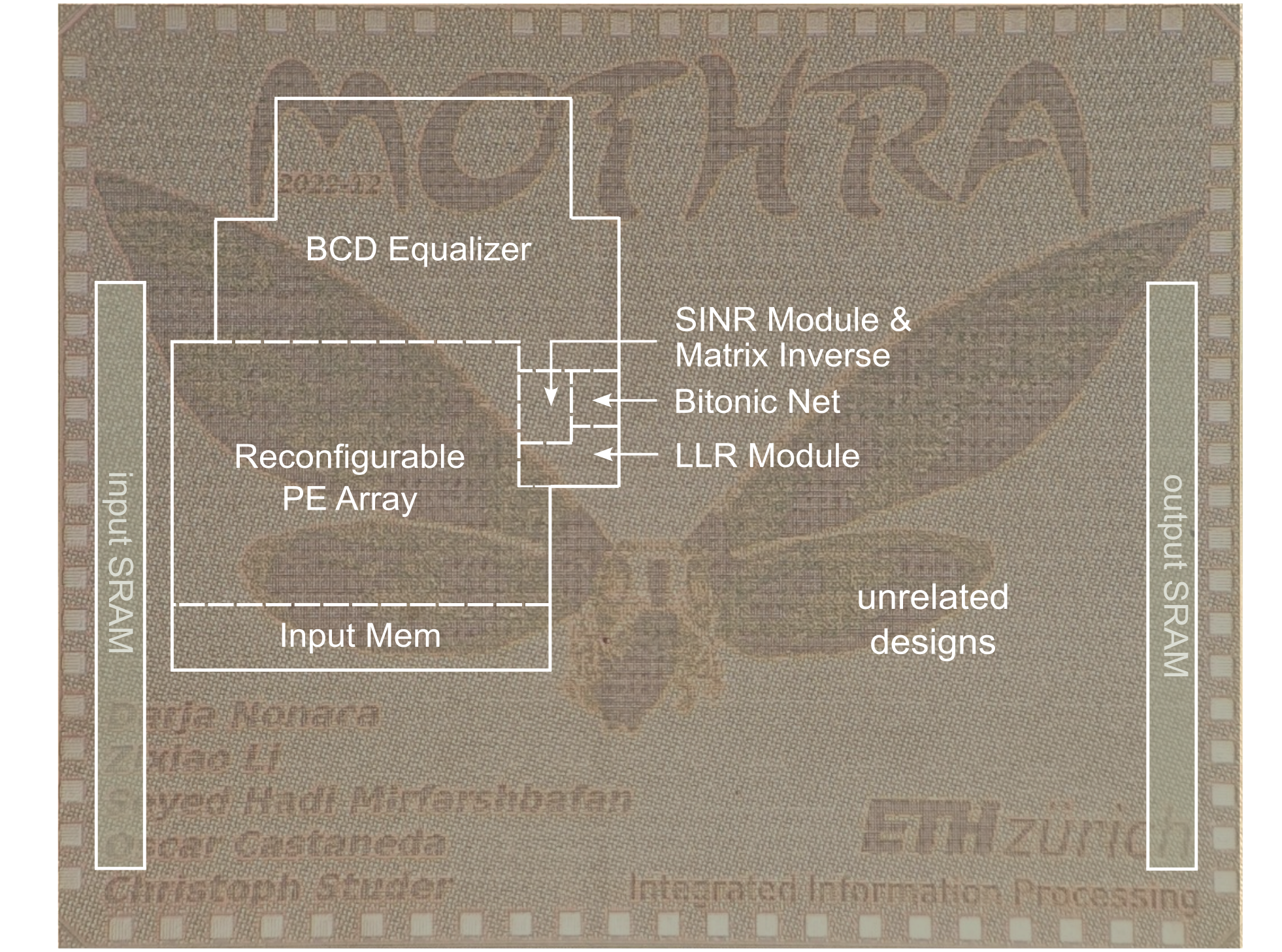}
\caption{Chip photo of the fabricated ASIC showing the GBCD core and its main modules. (There are unrelated designs sharing the same chip area.)}
\label{fig:floorplan}
\end{figure}

\fref{fig:floorplan} shows a micrograph of the ASIC fabricated in $22$\,nm FD-SOI. The ASIC implements the proposed GBCD detector core, which is composed of the input memory, preprocessing, and BCD equalization modules shown in \fref{fig:top_level}. The ASIC also contains input and output SRAMs for testing purposes. Since the GBCD core includes its own input memory module, we exclude the test SRAMs from the area and power measurements. The GBCD core occupies a silicon area of $0.97$\,mm$^\textnormal{2}$ and supports a $128\times 16$ MIMO system with QPSK, 16-QAM, 64-QAM, and 256-QAM. 
We note that the fabricated chip includes other designs as well. Hence, the GBCD detector occupies only a small portion of the total die.

\subsection{ASIC Measurements} \label{sec:ASICMeasure}
As detailed in \fref{eq:throughput}, the throughput $\Theta$ depends on the number of transmissions $T$ per coherence block. Furthermore, \fref{eq:BCDutilization} indicates that $T$ affects the equalizer utilization $\eta_\text{equ}$, which in turn influences the average power $P$. 
We emphasize that for wideband mmWave communication systems, the value of~$T$ could be of the order of $10^3$. For example, consider an OFDM system with $2048$ subcarriers communicating over a bandwidth of $500$ MHz, which requires a minimum sampling rate of $1$\,G~samples per second. Assuming a coherence time of $0.1$\,ms and coherence bandwidth of $10$\,MHz (conservative values taken from  \cite{RailwayMMWave}), $T$ would be larger than $1000$.
Therefore, our goal is to explore the impact of $T$ on the power consumption and energy efficiency, rather than just measuring the design at a particular~$T$ value. However, due to the storage limit of the input SRAM, we are unable to measure the design when $T$ is larger than $54$. To make a comprehensive analysis and derive the asymptotic power $P(T \to \infty)$, we (i) build a theoretical model with unknown power parameters; (ii) measure the power at distinct values of $T$; (iii) fit the model; and (iv) check the R-square statistics to validate the fitting.

Specifically, we first model the relationship between total power $P$ and $T$ according to module utilization depicted in \fref{fig:time_schedule} and \fref{eq:BCDutilization}: 
\begin{align} \label{eq:powerbreak}
	P(T) = P_\text{pre} + \eta_\text{equ}P_\text{equ}+P_\text{sta} = \tilde{P}+\frac{T}{T+9}P_\text{equ},
\end{align}
where $P_\text{pre}$ and $P_\text{equ}$ stand for the dynamic power of the preprocessor and BCD equalizer (cf. \fref{fig:top_level}) at 100\% utilization, respectively. $P_\text{sta}$ represents the static power and we define $\tilde{P}=P_\text{pre}+P_\text{sta}$. We emphasize that the MF power is included in $P_{\text{pre}}$ as it is computed by the reconfigurable PE array in the preprocessor. 
Subsequently, we measure the total power $P$ at nine distinct values of $T$, ranging from $6$ to $54$ at equal intervals, and fit the values of $\tilde{P}$ and $P_\text{equ}$ with a nonlinear least-square solver.

\begin{figure}[t]
	\centering
	\includegraphics[width=1.0\columnwidth]{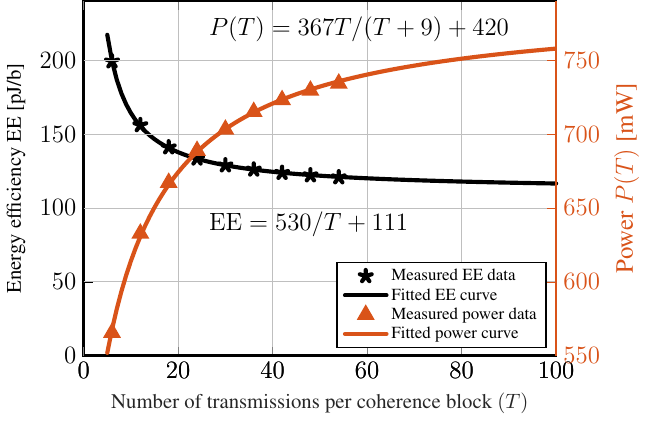}
	\caption{Energy efficiency and power vs. number $T$ of transmissions within a channel coherence block at VDD\,$=$\,0.8\,V and $f_\text{clk} =$\,887\,MHz.}
	\label{fig:EEversusT}
\end{figure}

\begin{figure}[t]
	\centering
	\includegraphics[width=1.0\columnwidth]{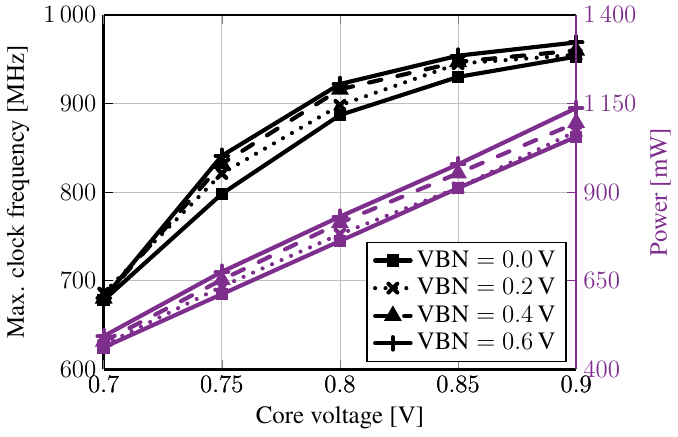}
	\caption{Frequency measurements (black lines) and power measurements (purple lines) for different core voltages and body biases at $T=54$. Lines with different marker shapes correspond to measurements with different forward body biases.}
	\label{fig:vol_bb_sweep}
\end{figure}

\begin{table*}[t]
\centering
\renewcommand{\arraystretch}{1.1}
\begin{minipage}[c]{1\textwidth}
\vspace{-0.1cm}
    \centering
    \caption{Comparison with state-of-the-art massive MU-MIMO data detectors}
       \label{tbl:implresults}
  \begin{threeparttable}
  \begin{tabular}{@{}l|>{\centering\arraybackslash}p{1cm}|>{\centering\arraybackslash}p{1cm}|>
  {\centering\arraybackslash}p{1cm}|>{\centering\arraybackslash}p{1cm}|>
  {\centering\arraybackslash}p{1cm}|>
  {\centering\arraybackslash}p{0.7cm}|>
  {\centering\arraybackslash}p{0.7cm}|>
  {\centering\arraybackslash}p{0.7cm}|>
  {\centering\arraybackslash}p{0.7cm}@{}}
    \hline
  \toprule
    & Prabhu & Jeon & Tang & Tang & Peng & \multicolumn{2}{c|}{Liu} & \multicolumn{2}{c}{This work}\\
    & \cite{prabhu17} & \cite{jeon19b} & \cite{tang18} & \cite{tang19} & \cite{peng18b} & \multicolumn{2}{c|}{\cite{liu20}} & \multicolumn{2}{c}{}\\
  \midrule
  {Algorithm} & CHD & LAMA & EPD & MPD & WeJi & \multicolumn{2}{c|}{RCG} & \multicolumn{2}{c}{GBCD} \\
  {MIMO size ($B\times U$)}  & $128\times 8$ & $256\times 32$ & $128\times 16$ & $128\times 32$ & $128\times 8$ & \multicolumn{2}{c|}{$128\times 8$} & \multicolumn{2}{c}{$128\times 16$}\\
  {Modulation [QAM]} & $256$ & $256$ & $4$ to $256$& $256$ & $64$ & \multicolumn{2}{c|}{$64$} & \multicolumn{2}{c}{$4$ to $256$} \\
  {Soft output} & No & Yes & No & No & Yes & \multicolumn{2}{c|}{Yes} & \multicolumn{2}{c}{Yes} \\  
  {Realistic channels} & Yes & Yes & Yes & No\tnote{a} & {No}\tnote{b} & \multicolumn{2}{c|}{{No}\tnote{b}} & \multicolumn{2}{c}{Yes}\\
  \midrule
  {Technology [nm]} & 28 & 28 & 28 & 40 & 65 & \multicolumn{2}{c|}{65} & \multicolumn{2}{c}{22} \\
  {Core voltage [V]} & 0.9 & 0.9 & 1.0 & 0.9 & 1.0 &\multicolumn{2}{c|}{1.2} & \multicolumn{2}{c}{0.8} \\
  Max.\ frequency [MHz] & 300 & 400 & 569 & 425 & 680 &\multicolumn{2}{c|}{500} & \multicolumn{2}{c}{887} \\
  Throughput [Gbps] & 0.3 & 0.35 & 1.8 & 2.76 & 1.02 &\multicolumn{2}{c|}{1.5} & \multicolumn{2}{c}{7.1} \\
    Norm. throughput\tnote{c}~ [Gbps] & 0.38 & 0.45 & 2.29 & 5.02 & 3.01 &\multicolumn{2}{c|}{4.43} & \multicolumn{2}{c}{7.1} \\
  \midrule 
  {Gram \& MF measured} & No\tnote{d} & No & No & No & Yes & Yes & No & Yes & No \\
  Core area [$\mathrm{mm^2}$] & -- & 0.37 & 2.0 & 0.58 & 2.57 & 3.50 & 1.62 & 0.97 & 0.43 \\
  Power [mW] & 18 & 151 & 127 & 221 & 650 & 557 & 120 & 787 & 367 \\
  Area eff.\tnote{c} \tnote{e} \tnote{f}~ [$\mathrm{Gbps/mm^2}$] & -- & 3.90 & 1.86 & 57.20 & 5.12 & 5.53 & 11.94 & 7.32 & 16.51 \\
  Energy eff.\tnote{c} \tnote{e} \tnote{f}~ [pJ/b] & 74.50 & 133.92 & 35.48 & 17.40 & 276.08 & 111.72 & 24.07 & 110.85 & 51.69
  \tabularnewline  
  \bottomrule
  \end{tabular}
  \vspace{0.1cm}
  \begin{tablenotes}
      \footnotesize
      \item[a] Simulations in \cite{tang19} show that MPD suffers from severe performance degradation in realistic correlated channels.
      \item[b] Realistic channel support requires more iterations than those reported in the original work \cite{peng18b}.
      \item[c] Technology normalized to $22$\,nm at $0.8$\,V nominal core voltage, given the assumptions: $f_\text{clk}\sim s$, $\mathrm{Area}\sim 1/s^2$ and $\mathrm{Power}\sim 1/(V^2)$, where $s$ is the ratio of technology nodes and $V$ is the ratio of core voltages \cite{Rabaey_book}. 
      \item[d] CHD implements the Gram and MF on chip, but the power measurement excludes these operations.
      \item[e] UE number is normalized to 16, assuming that area and power scale with $(U/16)^2$, and throughput also scales with $U/16$.
      \item[f] Area and energy efficiency are defined as throughput/area and power/throughput, respectively.
    \end{tablenotes}
  \end{threeparttable}
  \end{minipage}
  \end{table*}

\fref{fig:EEversusT} presents the fitted curve $P$ against $T$ (red line) based on the power measurements (red markers) at the maximum frequency of $887$\,MHz for the nominal core voltage of $0.8$\,V with zero body biasing. The R-squared statistic of the fitted curve is $R^2=0.9994$, close to the ideal $R^2=1$ for perfect fitting. This high R-squared value demonstrates the accuracy of our model in~\fref{eq:powerbreak}. After fitting, the equalizer power~$P_\text{equ}$ and asymptotic total power $P(T \to \infty)$ are $367$\,mW and $787$\,mW, respectively. Based on \fref{eq:throughput}, the asymptotic throughput at $887$\,MHz maximum clock frequency is $7.1$\,Gbps, which equals to the throughput without considering preprocessing. We further compute the energy efficiency as $\text{EE}=P(T)/\Theta(T)$ by plugging in \fref{eq:throughput} and \fref{eq:powerbreak}, and present the EE versus $T$ curve (black line) in \fref{fig:EEversusT}. 
As expected, the EE improves as~$T$ increases.
The asymptotic EE of the
entire design and the BCD equalizer with 256-QAM are 111\,pJ/b and 51.69\,pJ/b, respectively.

The frequency and power measurements for different core voltages and body biases at room temperature are shown in \fref{fig:vol_bb_sweep}, with the power measured at $T=54$. At a nominal supply of $0.8$\,V without body bias, the GBCD core runs at a maximum clock frequency of $887$\,MHz. 
By increasing the core voltage to $0.9$\,V and applying a forward body bias (FBB) of $0.4$\,V, the clock frequency reaches $969$\,MHz, which corresponds to a $9.2$\% throughput increase. For low-power applications, one can decrease the power consumption by $40$\%  from $761$\,mW to $461$\,mW at $680$\,MHz clock frequency by reducing the core voltage to $0.7$\,V and applying zero body bias.

\subsection{Comparison with Other Designs}

\fref{tbl:implresults} compares GBCD with the fabricated state-of-the-art massive MU-MIMO data detectors. Some designs, such as~\cite{jeon19b, tang18, tang19}, do not include the Gram and MF modules and take~$\bG$ and~$\bmy^{\mathrm{MF}}$ as inputs. To enable a fair comparison, we also provide measurement results without the Gram and MF modules. GBCD's main advantages compared to the state-of-the-art designs are (i) superior BLER performance under realistic channel conditions as demonstrated in \fref{fig:BLER}, (ii)~$1.4\times$ to $18.6\times$ higher normalized throughput, and (iii)~$1.3\times$ to $8.9\times$ higher area efficiency (except for the MPD detector~\cite{tang19}, which suffers large performance degradation under realistic channels according to their results).

The normalized area efficiency of GBCD is $1.4 \times$ better than WeJi~\cite{peng18b} and $1.3\times$ better than RCG~\cite{liu20}.
In terms of normalized energy efficiency, GBCD outperforms the LMMSE detector (CHD)~\cite{prabhu17} and WeJi by $1.3\times$ and $2.5\times$, respectively, but is outperformed by RCG by $2.1\times$ if Gram and MF computations are excluded. 
We note that the higher energy efficiency of RCG is partly achieved by limiting the number of iterations to two. Note, however, that \fref{fig:BLER} demonstrates that two iterations with RCG incurs a prohibitively high error floor for mmWave channels. To mitigate this issue, more than six RCG iterations are needed, which also approximately triples  the reported equalization power consumption of RCG in~\cite{liu20}.

Both LAMA~\cite{jeon19b} and EPD~\cite{tang18} demonstrate good error-rate performance with realistic channels. However, to achieve this, LAMA suffers from low throughput and EPD from large silicon area. Specifically, the area efficiency of LAMA and EPD are $4.2 \times$ and $8.9 \times$ lower than that of GBCD. We note that EPD exhibits a $1.5 \times$ energy efficiency advantage over GBCD, but its large area poses a challenge for deployment in OFDM systems, which need small data-detector cores to process the large number of parallel subcarriers. Moreover, EPD only generates hard outputs, which makes it far less effective in real-world systems that use forward error correction.

While our design exhibits inferior area and energy efficiency than the MPD detector~\cite{tang19}, simulation results from~\cite{tang19} reveal that MPD suffers a high error floor in correlated mmWave channels. In contrast, the proposed GBCD detector has demonstrated its effectiveness in both mmWave non-LoS and LoS channels in \fref{sec:performance}. Therefore, for practical scenarios, the GBCD detector is to be preferred.

\section{Conclusions}
\label{sec:conclusion}

We have proposed a deep-unfolding-assisted GBCD data detector suitable for the massive MU-MIMO-OFDM uplink, which is robust to correlated mmWave channels and enables an efficient VLSI implementation. Simulations with mmWave channels have shown that the error-rate performance of the proposed GBCD-PME is at least on-par with LMMSE equalization when the ratio between BS antenna number and UE number is large. The GBCD-PME also offers $4$\,dB to $10$\,dB SNR gain at $1\%$~BLER compared to LMMSE equalization when the number of UE and BS antennas are comparable. Furthermore, GBCD-PME (often significantly) outperforms all reference state-of-the-art designs in terms of error-rate performance.

We have developed an area-efficient VLSI architecture for GBCD, which can perform both BOX and PME denoising and supports QPSK to 256-QAM modulation within the same circuitry.
We have fabricated an ASIC implementing the proposed method for a system with $128$ BS antennas and $16$ UEs in $22$\,nm FD-SOI. With the proposed reconfigurable PE array and PME approximation, the ASIC achieves a throughput of $7.1$\,Gbps and occupies less than $1$\,mm$^\textnormal{2}$ including preprocessing circuitry. Compared to fabricated data  detectors which can deal with correlated mmWave channels, our design provides $1.3\times$ to $8.9\times$ higher area efficiency. 

The small silicon footprint, high throughput, and low preprocessing latency, make GBCD the favorable choice for multi-antenna OFDM systems, which require a large number of parallel data-detector cores and significant data buffering for a large number of subcarriers.

\section*{Acknowledgments}
The work of SHM, OC, and CS was supported in part by the Swiss National Science Foundation, an ETH Zurich grant, and has received funding from the Swiss State Secretariat for Education, Research, and Innovation (SERI) under the SwissChips initiative.
The authors would like to thank GlobalFoundries for providing silicon fabrication through the 22FDX university program. The authors also thank Darja Nonaca for her contributions to the Mothra~ASIC.



\balance

\end{document}